%% file: _main.tex
\documentclass[reqno,letterpaper]{amsart}
\usepackage{import}
\usepackage{url}
\usepackage{tikz, pgf}
\usetikzlibrary{calc}
\usetikzlibrary{decorations.pathreplacing}
\usepackage{amsmath, amsthm, bm, bbm, amssymb}
\usepackage{mathtools}

\usepackage{amsfonts}
\newenvironment{proofof}[1]{\renewcommand{\proofname}{Proof of #1}\proof}{\endproof}
\usepackage{algpseudocode, algorithm}
\usepackage{graphicx}
\usepackage{datetime}
\usepackage{comment}
\usepackage{multirow}
\usepackage{wrapfig}
\usepackage{graphicx} %
\usepackage{tablefootnote}
\usepackage[numbers,sort&compress]{natbib}
\usepackage[colorlinks,citecolor=blue,urlcolor=black,linkcolor=blue,pdfborder={0 0 0}]{hyperref}

\usepackage[capitalize,sort]{cleveref}  

\crefname{nlem}{Lemma}{Lemmas}
\crefname{nprop}{Proposition}{Propositions}
\crefname{ncor}{Corollary}{Corollaries}
\crefname{nthm}{Theorem}{Theorems}
\crefname{assumption}{Assumption}{Assumptions}

\usepackage{color}

\usepackage[arxiv]{optional}

\usepackage{jhh-misc2}

\algrenewcommand{\algorithmiccomment}[1]{\hfill$\triangleright$ #1}

\usepackage{caption}
\usepackage{subcaption}

\graphicspath{ {figures-updated/} }
\allowdisplaybreaks[3]

\input{shortcuts}
\usepackage{autonum}

\makeatletter
\newcommand\footnoteref[1]{\protected@xdef\@thefnmark{\ref{#1}}\@footnotemark}
\makeatother

\numberwithin{equation}{section}

\begin{document}

\title[Coresets for Scalable Bayesian Logistic Regression]{Coresets for Scalable \\ Bayesian Logistic Regression}

\author[J.~H.~Huggins]{Jonathan H.~Huggins}
\address{Computer Science and Artificial Intelligence Laboratory (CSAIL) \\ Massachusetts Institute of Technology}
\urladdr{http://www.jhhuggins.org/}
\email{jhuggins@mit.edu}

\author[T.~Campbell]{Trevor Campbell}
\address{Computer Science and Artificial Intelligence Laboratory (CSAIL) \\ Massachusetts Institute of Technology}
\urladdr{http://www.trevorcampbell.me/}
\email{tdjc@mit.edu}

\author[T.~Broderick]{Tamara Broderick}
\address{Computer Science and Artificial Intelligence Laboratory (CSAIL) \\ Massachusetts Institute of Technology}
\urladdr{http://www.tamarabroderick.com}
\email{tbroderick@csail.mit.edu}

\date{\today}

\begin{abstract}
\input{abstract.tex}
\end{abstract}

\maketitle

\input{content.tex}

\subsubsection*{Acknowledgments}
All authors are supported by the Office of Naval Research under ONR MURI grant N000141110688.
JHH is supported by the U.S. Government under FA9550-11-C-0028 and awarded by the DoD, Air Force
Office of Scientific Research, National Defense Science and Engineering Graduate
(NDSEG) Fellowship, 32 CFR 168a.

\input{appendix.tex}

\bibliographystyle{abbrvnat}
\bibliography{library}

\end{document}

%% file: shortcuts.tex
\newcommand{\ranges}{\textbf{ranges}}
\newcommand{\range}{\textbf{range}}
\newcommand{\ball}{\mathbb{B}}
\newcommand{\sphere}{\mathbb{S}}

\newcommand{\dsname}[1]{\textsc{#1}\xspace}
\newcommand{\binDat}{\dsname{Binary5}}
\newcommand{\BinDat}{\dsname{Binary10}}
\newcommand{\mixture}{\dsname{Mixture}}
\newcommand{\webspam}{\dsname{Webspam}}
\newcommand{\covtype}{\dsname{CovType}}
\newcommand{\chemre}{\dsname{ChemReact}}

%% file: abstract.tex
The use of Bayesian methods in large-scale data settings is attractive because of 
the rich hierarchical models, uncertainty quantification, and prior specification they provide. 
Standard Bayesian inference algorithms are computationally expensive, however, making 
their direct application to large datasets difficult or infeasible. 
Recent work on scaling Bayesian inference has focused on modifying the underlying algorithms 
to, for example, use only a random data subsample at each iteration. 
We leverage the insight that data is often redundant to instead obtain a weighted subset 
of the data (called a \emph{coreset}) that is much smaller than the original dataset. 
We can then use this small coreset in any number of existing posterior inference algorithms without modification. 
In this paper, we develop an efficient coreset construction algorithm for Bayesian logistic regression models. 
We provide theoretical guarantees on the size and approximation quality of the coreset -- both for fixed, known datasets, 
and in expectation for a wide class of data generative models. 
Crucially, the proposed approach also permits efficient construction of the coreset in both 
streaming and parallel settings, with minimal additional effort. 
We demonstrate the efficacy of our approach on a number of synthetic and real-world 
datasets, and find that, in practice, the size of the coreset is 
independent of the original dataset size. 
Furthermore, constructing the coreset takes a negligible amount of 
time compared to that required to run MCMC on it.

%% file: content.tex
\newcommand{\suppmat}{\opt{nips}{Appendix}\opt{arxiv}{Appendix}}

\section{Introduction}

Large-scale datasets, comprising tens or hundreds of millions of observations, are becoming the norm 
in scientific and commercial applications ranging from population genetics to advertising.
At such scales even simple operations, such as examining each data point a small number of times, 
become burdensome; it is sometimes not possible to fit all data in the physical memory of a single machine.
These constraints have, in the past, limited practitioners to relatively simple statistical modeling approaches.
However, the rich hierarchical models, uncertainty quantification, and prior specification provided
by Bayesian methods have motivated substantial recent effort in making Bayesian
inference procedures, which are often computationally expensive, scale to the large-data setting. 

The standard approach to Bayesian inference for large-scale data is to modify a specific inference algorithm,
such as MCMC or variational Bayes, to handle distributed or streaming processing of data.
Examples include subsampling and streaming methods for 
variational Bayes~\citep{Hoffman:2013,Broderick:2013b,Campbell:2015}, subsampling methods for 
MCMC~\opt{arxiv}{\citep{Welling:2011,Ahn:2012,Bardenet:2014,Korattikara:2014,Maclaurin:2014,Bardenet:2015}}\opt{nips}{\citep{Welling:2011,Maclaurin:2014,Bardenet:2015}}, and 
distributed ``consensus'' methods for MCMC~\citep{Scott:2013,Srivastava:2015,Rabinovich:2015,Entezari:2016}. 
Existing methods, however, suffer from both practical and theoretical limitations. 
Stochastic variational inference~\citep{Hoffman:2013} and subsampling MCMC methods use a new random
subset of the data at each iteration, which requires random 
access to the data and hence is infeasible for very large datasets that do not fit into memory.
Furthermore, in practice, subsampling MCMC methods have been found 
to require examining a constant fraction of the data at each iteration, severely limiting the computational gains 
obtained~\opt{arxiv}{\citep{Pillai:2014,Bardenet:2015,Teh:2016,Betancourt:2015,Alquier:2016}}\opt{nips}{\citep{Teh:2016,Betancourt:2015}}.
More scalable methods such as consensus MCMC~\opt{arxiv}{\citep{Scott:2013,Srivastava:2015,Rabinovich:2015,Entezari:2016}}\opt{nips}{\citep{Scott:2013,Srivastava:2015,Rabinovich:2015}} and streaming
variational Bayes~\citep{Broderick:2013b,Campbell:2015} lead to gains in computational efficiency, but lack 
rigorous justification and provide no guarantees on the quality of inference.

An important insight in the large-scale setting is that much of the data is often \emph{redundant}, though there 
may also be a small set of data points that are distinctive. 
For example, in a large document corpus, one news article about a hockey game may serve as an excellent representative of
hundreds or thousands of other similar pieces about hockey games. 
However, there may only be a few articles about luge, so it is also important to include at least one article about luge. 
Similarly, one individual's genetic information may serve as a strong representative of other individuals from the same 
ancestral population admixture, though some individuals may be genetic outliers. 
We leverage data redundancy to develop a scalable Bayesian inference framework that modifies
the \emph{dataset} instead of the common practice of modifying the inference algorithm.
Our method, which can be thought of as a preprocessing step, 
constructs a \emph{coreset} -- a small, weighted subset of the data that approximates
the full dataset~\citep{Agarwal:2005,Feldman:2011a} -- that can be used in many standard inference procedures
to provide posterior approximations with guaranteed quality. 
The scalability of posterior inference with a coreset thus simply depends on the coreset's growth 
with the full dataset size.  
To the best of our knowledge, coresets have not previously been used in a Bayesian setting.

The concept of coresets originated in computational geometry (e.g.~\citep{Agarwal:2005}), but then
became popular in theoretical computer science as a way to efficiently solve clustering problems such as 
$k$-means and PCA~(see \citep{Feldman:2011a,Feldman:2013} and references therein). 
Coreset research in the machine learning community has focused on scalable clustering in the optimization
setting~\opt{arxiv}{\citep{Feldman:2013,Bachem:2015,Lucic:2016,Bachem:2016}}\opt{nips}{\citep{Lucic:2016,Bachem:2016}},
with the exception of \citet{Feldman:2011b}, who developed a coreset algorithm for Gaussian mixture models. 
Coreset-like ideas have previously been explored for maximum likelihood-learning of logistic regression models,
though these methods either lack rigorous justification or have only asymptotic guarantees~(see~\citep{Han:2016} 
and references therein\opt{arxiv}{ as well as \citep{Madigan:2002}, which develops a methodology applicable beyond logistic regression}). 

The job of the coreset in the Bayesian setting is to provide an approximation
of the full data log-likelihood up to a multiplicative error uniformly over the parameter space. 
As this paper is the first foray into applying coresets in Bayesian inference, we begin
with a theoretical analysis of the quality of the posterior distribution obtained from
such an approximate log-likelihood.
The remainder of the paper develops the efficient construction of small coresets for Bayesian logistic regression, 
a useful and widely-used model for the ubiquitous problem of binary classification.
We develop a coreset construction algorithm, the output of which uniformly approximates
the full data log-likelihood over parameter values in a ball with a user-specified radius.
The approximation guarantee holds for a given dataset with high probability.
We also obtain results showing that the boundedness of the parameter space 
is necessary for the construction of a nontrivial coreset, as well as results characterizing
the algorithm's expected performance under a wide class of data-generating distributions. 
Our proposed algorithm is applicable in both the streaming and distributed computation settings, 
and the coreset can then be used by any inference algorithm which accesses the (gradient of the) log-likelihood 
as a black box. Although our coreset algorithm is specifically for logistic regression, our approach 
is broadly applicable to other Bayesian generative models.

Experiments on a variety of synthetic and real-world datasets validate our approach and demonstrate 
robustness to the choice of algorithm hyperparameters.
An empirical comparison to random subsampling shows that, in many cases, coreset-based posteriors are 
orders of magnitude better in terms of maximum mean discrepancy, including on a challenging
100-dimensional real-world dataset. 
Crucially, our coreset construction algorithm adds negligible computational overhead to the inference
procedure. 
All proofs are deferred to the \suppmat. 

\section{Problem Setting}

We begin with the general problem of Bayesian posterior inference.
Let $\mcD = \{(X_{n}, Y_{n})\}_{n=1}^{N}$ be a dataset, where $X_{n} \in \mcX$ is a vector of
covariates and $Y_n \in \mcY$ is an observation.
Let $\pi_{0}(\theta)$ be a prior density on a parameter $\theta \in \Theta$ %
and let $p(Y_n \given X_n, \theta)$ be the likelihood of observation $n$ given the parameter $\theta$.
\opt{arxiv}{
The Bayesian posterior is given by the density
\[
\pi_{N}(\theta) &\defined \frac{\exp(\mcL_{N}(\theta))\pi_{0}(\theta)}{\mcE_{N}},
\]
where
$
\mcL_{N}(\theta) \defined \sum_{n=1}^{N} \ln p(Y_{n} \given X_{n}, \theta)
$
is the model log-likelihood and
\[
\mcE_{N} \defined \int\exp(\mcL_{N}(\theta))\pi_{0}(\theta)\,\dee \theta
\]
is the marginal likelihood (a.k.a.~the model evidence). 
}\opt{nips}{
The Bayesian posterior is given by the density $\pi_N(\theta)$, where
\[
\pi_{N}(\theta) &\defined \frac{\exp(\mcL_{N}(\theta))\pi_{0}(\theta)}{\mcE_{N}}, && 
\mcL_{N}(\theta) \defined \sum_{n=1}^{N} \ln p(Y_{n} \given X_{n}, \theta), &&
\mcE_{N} \defined \int\exp(\mcL_{N}(\theta))\pi_{0}(\theta)\,\dee \theta.
\]
}
Our aim is to construct a weighted dataset $\tilde\mcD = \{(\gamma_{m}, \tX_{m}, \tY_{m})\}_{m=1}^{M}$
with $M \ll N$ such that the weighted log-likelihood
$
\tilde\mcL_{N}(\theta) = \sum_{m=1}^{M} \gamma_{m} \ln p(\tY_{n} \given \tX_{m}, \theta)
$
satisfies
\[
|\mcL_{N}(\theta) - \tilde\mcL_{N}(\theta)| &\le \varepsilon|\mcL_{N}(\theta)|, 
\quad \forall \theta \in \Theta. \label{eq:lik-multi-error}
\]
If $\tilde\mcD$ satisfies \cref{eq:lik-multi-error}, it is called an
\emph{$\varepsilon$-coreset of $\mcD$}, and the approximate posterior
\[
\tpi_{N}(\theta) &= \frac{\exp(\tilde\mcL_{N}(\theta))\pi_{0}(\theta)}{\tilde \mcE_{N}}, & \text{with} \quad 
\tilde \mcE_{N} &= \int\exp(\tilde\mcL_{N}(\theta))\pi_{0}(\theta)\,\dee \theta,
\]
has a marginal likelihood $\tilde \mcE_{N}$ which approximates the true
marginal likelihood $\mcE_{N}$, shown by \cref{prop:marginal-likelihood-approx}.
Thus, from a Bayesian perspective, the $\varepsilon$-coreset is a useful notion of 
approximation. 

\bnprop\label{prop:marginal-likelihood-approx}
Let $\mcL(\theta)$ and $\tilde\mcL(\theta)$ be arbitrary non-positive log-likelihood functions that
satisfy $|\mcL(\theta) - \tilde\mcL(\theta)| \le \varepsilon|\mcL(\theta)|$ for all $\theta \in \Theta$.
Then for any prior $\pi_{0}(\theta)$ such that the marginal likelihoods
\[
\mcE = \int\exp(\mcL(\theta))\pi_{0}(\theta)\,\dee \theta
\qquad \text{and} \qquad
\tilde\mcE = \int\exp(\tilde\mcL(\theta))\pi_{0}(\theta)\,\dee \theta
\]
are finite, %
the marginal likelihoods satisfy
\opt{arxiv}{
\[
|\ln \mcE - \ln \tilde\mcE| \le \varepsilon|\ln \mcE|.
\]
}\opt{nips}{
$ |\ln \mcE - \ln \tilde\mcE| \le \varepsilon|\ln \mcE|$.
}
\enprop

\section{Coresets for Logistic Regression}
\begin{algorithm}[t]
\caption{Construction of logistic regression coreset}\label{alg:lr-coreset}
\begin{algorithmic}[1] 
\Require Data $\mcD$, $k$-clustering $\mcQ$, radius $R > 0$, tolerance $\varepsilon > 0$, failure rate $\delta \in (0,1)$
\For{$n=1,\dots,N$} \Comment{calculate sensitivity upper bounds using the $k$-clustering}
\State $m_{n} \gets \frac{N}{1 + \sum_{i=1}^{k}|G_{i}^{(-n)}|e^{-R\|\bZ_{G,i}^{(-n)} - Z_{n}\|_{2}}}$
\EndFor
\State $\bbm_N \gets \frac{1}{N}\sum_{n=1}^{N}m_{n}$
\State $M \gets \left\lceil\frac{c \bbm_N}{\varepsilon^{2}}[(D + 1)\log \bbm_{N} + \log(1/\delta)]\right\rceil$ \Comment{coreset size; $c$ is from proof of \cref{thm:making-coresets}}
\For{$n=1,\dots,N$}
\State $p_{n} \gets \frac{m_{n}}{N\bbm_N}$ \Comment{importance weights of data}
\EndFor
\State $(K_{1},\dots,K_{N}) \dist \distMulti(M, (p_{n})_{n=1}^{N})$ \Comment{sample data for coreset}
\For{$n=1,\dots,N$} \Comment{calculate coreset weights}
\State $\gamma_{n} \gets \frac{K_{n}}{p_{n}M}$
\EndFor
\State $\tilde\mcD \gets \{(\gamma_{n}, X_{n}, Y_{n}) \given \gamma_{n} > 0 \}$  \Comment{only keep data points with non-zero weights}
\State \Return $\tilde\mcD$
\end{algorithmic}
\end{algorithm}

\subsection{Coreset Construction}

In logistic regression, the covariates are real feature vectors $X_n \in \reals^D$,
the observations are labels $Y_n \in \{-1, 1\}$, $\Theta \subseteq \reals^D$, 
and the likelihood is defined as
\[
p(Y_n \given X_n, \theta) 
&= p_{logistic}(Y_{n} \given X_{n}, \theta)
\defined \frac{1}{1+\exp\left(-Y_n X_n\cdot \theta\right)}.
\]
The analysis in this work allows any prior $\pi_0(\theta)$; common choices are
the Gaussian, Cauchy~\citep{Gelman:2008}, and spike-and-slab~\opt{arxiv}{\citep{Mitchell:1988,George:1993}}\opt{nips}{\citep{George:1993}}.
For notational brevity, we define $Z_n \defined Y_nX_n$, and let
$\phi(s) \defined \ln(1+\exp(-s))$.
Choosing the optimal $\eps$-coreset is not computationally feasible, so we take 
a less direct approach. 
We design our coreset construction algorithm and prove its correctness
using a quantity $\sigma_n(\Theta)$ called the \emph{sensitivity} \cite{Feldman:2011a},
which quantifies the redundancy of a particular data point $n$ -- the larger the
sensitivity, the less redundant. 
In the setting of logistic regression, we have that the sensitivity is
\[
\sigma_{n}(\Theta) 
\defined \sup_{\theta \in \Theta} \frac{N\,\phi(Z_{n} \cdot \theta)}{\sum_{\ell=1}^{N} \phi(Z_{\ell} \cdot \theta)}.
\]
Intuitively, $\sigma_n(\Theta)$ captures how much influence data point $n$ has on
the log-likelihood $\mcL_N(\theta)$ when varying the parameter $\theta \in \Theta$, 
and thus data points with high sensitivity should be included in the coreset.
Evaluating $\sigma_n(\Theta)$ exactly is not tractable, however,
so an upper bound $m_n \geq \sigma_n(\Theta)$ must be used in its place.
Thus, the key challenge is to efficiently compute a tight upper bound on the
sensitivity.

For the moment we will consider $\Theta = \ball_{R}$ for any $R > 0$, where 
$\ball_{R} \defined \{ \theta \in \reals^{D} \given \|\theta\|_{2} \le R\}$;
we discuss the case of $\Theta = \reals^{D}$ shortly. 
Choosing the parameter space to be a Euclidean ball is reasonable since data
is usually preprocessed to have mean zero and variance 1 (or, for sparse data, to be between -1 and 1),
so each component of $\theta$ is typically in a range close to zero (e.g.~between -4 and 4)~\citep{Gelman:2008}. 

The idea behind our sensitivity upper bound construction is that we would expect data points 
that are bunched together to be redundant while data points that are far from from other data have
a large effect on inferences. 
Clustering is an effective way to summarize data and detect outliers, so we will
use a \emph{$k$-clustering} of the data $\mcD$ to construct the sensitivity bound. 
A $k$-clustering is given by $k$ cluster centers $\mcQ = \{Q_{1},\dots,Q_{k}\}$. 
Let $G_{i} \defined \{ Z_{n} \given i = \argmin_{j}\|Q_{j} - Z_{n}\|_{2} \}$ be the set of vectors closest to center $Q_{i}$
and let $G_{i}^{(-n)} \defined G_{i} \setminus \{Z_{n}\}$.
Define $Z_{G,i}^{(-n)}$ to be a uniform random vector from $G_{i}^{(-n)}$ and let 
$\bZ_{G,i}^{(-n)} \defined \EE[Z_{G,i}^{(-n)}]$ be its mean. 
The following lemma uses a $k$-clustering to establish an efficiently computable
upper bound on $\sigma_n(\ball_{R})$:

\bnlem \label{lem:sensitivity-upper-bound}
For any $k$-clustering $\mcQ$,
\[
\sigma_{n}(\ball_{R}) 
\le m_n \defined \frac{N}{1 + \sum_{i=1}^{k}|G_{i}^{(-n)}|e^{-R\|\bZ_{G,i}^{(-n)} - Z_{n}\|_{2}}}.
\label{eq:sensitivity-upper-bound}
\]
Furthermore, $m_{n}$ can be calculated in $O(k)$ time. 
\enlem

The bound in \cref{eq:sensitivity-upper-bound} captures the intuition that if the data forms tight 
clusters (that is, each $Z_{n}$ is close to one of the cluster centers), we expect each cluster 
to be well-represented by a small number of typical data points.
For example, if $Z_{n} \in G_{i}$, $\|\bZ_{G,i}^{(-n)} - Z_{n}\|_{2}$ is small, and 
$|G_{i}^{(-n)}| = \Theta(N)$, then $\sigma_{n}(\ball_{R}) = O(1)$. 
We use the (normalized) sensitivity bounds obtained from \cref{lem:sensitivity-upper-bound} 
to form an importance distribution $(p_{n})_{n=1}^{N}$ from which to sample the coreset.
If we sample $Z_{n}$, then we assign it weight $\gamma_{n}$ proportional to $1/p_{n}$. 
The size of the coreset depends on the mean sensitivity bound, 
the desired error $\varepsilon$, and a quantity closely related
to the VC dimension of $\theta \mapsto \phi(\theta \cdot Z)$, which we show is $D+1$. 
Combining these pieces we obtain \cref{alg:lr-coreset},
which constructs an $\varepsilon$-coreset with high probability by \cref{thm:coreset-algorithm}. 

\bnthm \label{thm:coreset-algorithm}
Fix $\varepsilon > 0$, $\delta \in (0,1)$, and $R > 0$. 
Consider a dataset $\mcD$ with $k$-clustering $\mcQ$. 
With probability at least $1 - \delta$, 
\cref{alg:lr-coreset} with inputs $(\mcD, \mcQ, R, \varepsilon, \delta)$
constructs an $\varepsilon$-coreset of $\mcD$ for logistic regression with parameter
space $\Theta = \ball_{R}$. 
Furthermore, \cref{alg:lr-coreset} runs in $O(Nk)$ time. 
\enthm

\bnrmk
The coreset algorithm is efficient with an $O(Nk)$ running time.
However, the algorithm requires a $k$-clustering, which must also be constructed. 
A high-quality clustering can be obtained cheaply via $k$-means++ in $O(Nk)$ time~\citep{Arthur:2007}, 
although a coreset algorithm could also be used. %
\enrmk

Examining \cref{alg:lr-coreset}, we see that the coreset size $M$ is of order
$\bbm_N \log \bbm_{N}$, where $\bbm_{N} = \frac{1}{N}\sum_{n} m_n$.
So for $M$ to be smaller than $N$, at a minimum, $\bbm_{N}$ should satisfy $\bbm_{N} = \tilde o(N)$,\footnote{Recall that the tilde notation suppresses logarithmic terms.}
and preferably $\bbm_{N} = O(1)$. 
Indeed, for the coreset size to be small, it is critical that 
(a) $\Theta$ is chosen such that most of the sensitivities satisfy $\sigma_{n}(\Theta) \ll N$ 
(since $N$ is the maximum possible sensitivity),
(b) each upper bound $m_{n}$ is close to $\sigma_{n}(\Theta)$, and
(c) ideally, that $\bbm_{N}$ is bounded by a constant. 
In \cref{sec:sensitivity-lower-bound}, we address (a) by providing sensitivity lower bounds, thereby showing 
that the constraint $\Theta = \ball_R$ is necessary for nontrivial sensitivities 
even for ``typical'' (i.e.~non-pathological) data.
We then apply our lower bounds to address (b) and show that our bound 
in \cref{lem:sensitivity-upper-bound} is nearly tight.
In \cref{sec:sensitivity-bound-performance}, we address (c) by establishing the expected performance of the
bound in \cref{lem:sensitivity-upper-bound} for a wide class of data-generating distributions.

\subsection{Sensitivity Lower Bounds}
\label{sec:sensitivity-lower-bound}

We now develop lower bounds on the sensitivity to demonstrate that essentially we 
must limit ourselves to bounded $\Theta$,\footnote{Certain pathological datasets allow us to use unbounded $\Theta$, but we do not assume we are given such data.} 
thus making our choice of $\Theta = \ball_{R}$
a natural one, and to show that the sensitivity upper bound from \cref{lem:sensitivity-upper-bound} 
is nearly tight.  

We begin by showing that in both the worst case and the average case, for all $n$, $\sigma_{n}(\reals^{D}) = N$, 
the maximum possible sensitivity -- even when the $Z_{n}$ are arbitrarily close.
Intuitively, the reason for the worst-case behavior is that if there is a separating hyperplane between
a data point $Z_{n}$ and the remaining data points, and $\theta$ is in the direction of that hyperplane, then
when $\|\theta\|_{2}$ becomes very large, $Z_{n}$ becomes arbitrarily more important than any other data point. 

\bnthm \label{thm:sensitivity-lower-bound}
For any $D \ge 3$, $N \in \nats$ and $0 < \eps' < 1$, there exists $\epsilon > 0$ and unit vectors 
$Z_{1},\dots,Z_{N} \in \reals^{D}$ such that for all pairs $n,n'$, $Z_{n} \cdot Z_{n'} \ge 1 - \eps'$
and for all $R > 0$ and $n$,
\[
\sigma_{n}(\ball_{R}) &\ge \frac{N}{1 + (N-1)e^{-R\eps\sqrt{\eps'}/4}}, &&\text{and hence} &\sigma_{n}(\reals^{D}) &= N.
\]
\enthm

The proof of \cref{thm:sensitivity-lower-bound} is based on choosing $N$ distinct unit vectors 
$V_{1},\dots,V_{N} \in \reals^{D-1}$ and setting 
$\eps = 1 - \max_{n \ne n'} V_{n} \cdot V_{n'} > 0$.
But what is a ``typical'' value for $\eps$? 
In the case of the vectors being uniformly distributed on the unit sphere,
we have the following scaling for $\eps$ as $N$ increases:
\bnprop \label{prop:epsilon-behavior}
If $V_{1},\dots,V_{N}$ are independent and uniformly distributed on the unit sphere
$\sphere^{D} \defined \{ v \in \reals^{D} \given \|v\| = 1 \}$ with $D \ge 2$, 
then with high probability
\[
1 - \max_{n \ne n'} V_{n} \cdot V_{n'} \ge C_{D} N^{-4/(D-1)},
\]
where $C_{D}$ is a constant depending only on $D$. 
\enprop

Furthermore, $N$ can be exponential in $D$ even with $\eps$ remaining very close
to 1:

\bnprop \label{prop:exponentially-many-distant-vectors}
For $N = \lfloor\exp((1-\eps)^{2}D/4)/\sqrt{2}\rfloor$, and
$V_{1},\dots,V_{N}$ \iid\ such that $V_{ni} = \pm \frac{1}{\sqrt{D}}$ with probability
$\half$, then with probability at least $\half$, 
$
1 - \max_{n \ne n'} V_{n} \cdot V_{n'} \ge \eps.
$
\enprop

\cref{prop:epsilon-behavior,prop:exponentially-many-distant-vectors} demonstrate that 
the data vectors $Z_{n}$ found in \cref{thm:sensitivity-lower-bound}
are, in two different senses, ``typical'' vectors and should not be thought of as
worst-case data only occurring in some ``negligible'' or zero-measure set.
These three results thus demonstrate that it is necessary to restrict attention to bounded $\Theta$. 
We can also use \cref{thm:sensitivity-lower-bound} to show that our sensitivity
upper bound is nearly tight. 

\bncor \label{prop:matching-sensitivity-bounds}
For the data $Z_{1},\dots,Z_{N}$ from \cref{thm:sensitivity-lower-bound},
\[
\frac{N}{1 + (N-1)e^{-R\eps\sqrt{\eps'}/4}} 
\le \sigma_{n}(\ball_{R}) 
\le \frac{N}{1 + (N-1)e^{-R\sqrt{2\eps'}}}.
\]
\encor

\subsection{$k$-Clustering Sensitivity Bound Performance} 
\label{sec:sensitivity-bound-performance}

While \cref{lem:sensitivity-upper-bound,prop:matching-sensitivity-bounds} provide an upper bound 
on the sensitivity given a fixed dataset, we would also like to understand how the expected mean
sensitivity increases with $N$.
We might expect it to be finite since the logistic regression likelihood model is parametric;
the coreset would thus be acting as a sort of approximate finite sufficient statistic. 
\cref{prop:mixture-upper-bound} characterizes the expected
performance of the upper bound from \cref{lem:sensitivity-upper-bound} under a wide class of 
generating distributions. This result demonstrates that, under reasonable conditions, the expected value
of $\bbm_{N}$ is bounded for all $N$. 
As a concrete example, \cref{cor:gaussian-upper-bound} specializes \cref{prop:mixture-upper-bound}
to data with a single shared Gaussian generating distribution.

\bnprop \label{prop:mixture-upper-bound}
Let $X_n \distind \distNorm(\mu_{L_{n}}, \Sigma_{L_{n}})$, where 
$L_n \distind \distMulti(\pi_{1}, \pi_{2}, \dots)$ is the mixture component responsible for
generating $X_n$.
For $n=1, \dots, N$, let $Y_n \in \{-1, 1\}$ be conditionally independent given $X_n$
and set $Z_n = Y_n X_n$. 
Select $0 < r < 1/2$, and define $\eta_i = \max(\pi_i - N^{-r}, 0)$. 
The clustering of the data implied by $(L_n)_{n=1}^N$ results in the expected sensitivity bound
\[
\EE\left[\bbm_N\right]
&\leq \frac{1}{N^{-1} + \sum_{i} \eta_i e^{-R\sqrt{A_iN^{-1}\eta_i^{-1} + B_{i}}}}
 + \sum_{i : \eta_i > 0}Ne^{-2 N^{1-2r}}
\overset{N \to \infty}{\to} \frac{1}{\sum_i \pi_i e^{-R\sqrt{B_i}}},
\]
where
\opt{arxiv}{
\[
A_i &\defined \tr\left[\Sigma_i\right] + \left(1-\by_i^2\right)\mu_i^T\mu_i, \\
B_i &\defined {\textstyle\sum_j} \pi_j\left(\tr\left[\Sigma_j\right] + \by_j^2 \mu_i^T\mu_i - 2\by_i\by_j \mu_i^T\mu_j + \mu_j^T\mu_j\right),
\]
and $\by_j = \EE\left[Y_1 | L_1 = j\right]$.
}%
\opt{nips}{
$A_i \defined \tr\left[\Sigma_i\right] + \left(1-\by_i^2\right)\mu_i^T\mu_i$, 
$B_i \defined {\textstyle\sum_j} \pi_j\left(\tr\left[\Sigma_j\right] + \by_j^2 \mu_i^T\mu_i - 2\by_i\by_j \mu_i^T\mu_j + \mu_j^T\mu_j\right)$,
and $\by_j = \EE\left[Y_1 | L_1 = j\right]$.}
\enprop
\bncor \label{cor:gaussian-upper-bound}
In the setting of \cref{prop:mixture-upper-bound}, if $\pi_{1} = 1$ and
all data is assigned to a single cluster, then there is a constant $C$ such that for sufficiently large $N$,
\opt{nips}{
$\EE\left[\bbm_N\right] \le C e^{R\sqrt{\tr\left[\Sigma_{1}\right] +(1- \by_1^2) \mu_{1}^T\mu_{1}}}$.
}%
\opt{arxiv}{
\[
\EE\left[\bbm_N\right] \le C e^{R\sqrt{\tr\left[\Sigma_{1}\right] +(1- \by_1^2) \mu_{1}^T\mu_{1}}}.
\]}
\encor

\subsection{Streaming and Parallel Settings}

\cref{alg:lr-coreset} is a batch algorithm, but it can easily be used in parallel 
and streaming computation settings using  standard methods from the coreset literature,
which are based on the following two observations (cf.~\citep[Section 3.2]{Feldman:2011b}):
\opt{nips}{\vspace{-.5em}}
\benum
\item If $\tilde\mcD_{i}$ is an $\varepsilon$-coreset for $\mcD_{i}$, $i=1,2$, then 
$\tilde\mcD_{1} \cup \tilde\mcD_{2}$ is an $\varepsilon$-coreset for $\mcD_{1} \cup \mcD_{2}$.
\item If $\tilde\mcD$ is an $\varepsilon$-coreset for $\mcD$ and $\tilde\mcD'$ is an  
$\varepsilon'$-coreset for $\tilde\mcD$, then $\tilde\mcD'$ is an 
$\varepsilon''$-coreset for $\mcD$, where $\varepsilon'' \defined (1 + \varepsilon)(1 + \varepsilon') - 1$. 
\eenum
\opt{nips}{\vspace{-.5em}}
We can use these observations to merge coresets that were constructed either
in parallel, or sequentially, in a binary tree. 
Coresets are computed for two data blocks, merged using observation 1, then compressed
further using observation 2. 
The next two data blocks have coresets computed and merged/compressed in the same manner, then the
coresets from blocks 1\&2 and 3\&4 can be merged/compressed analogously. 
We continue in this way and organize the merge/compress operations into a binary tree. Then, 
if there are $B$ data blocks total, only $\log B$ blocks ever need be maintained simultaneously.
In the streaming setting we would choose blocks of constant size, so $B = O(N)$,
while in the parallel setting $B$ would be the number of machines available.

\section{Experiments}

\begin{figure}[tb]
\begin{center}
\begin{subfigure}[b]{0.43\textwidth}
    \includegraphics[width=0.98\textwidth]{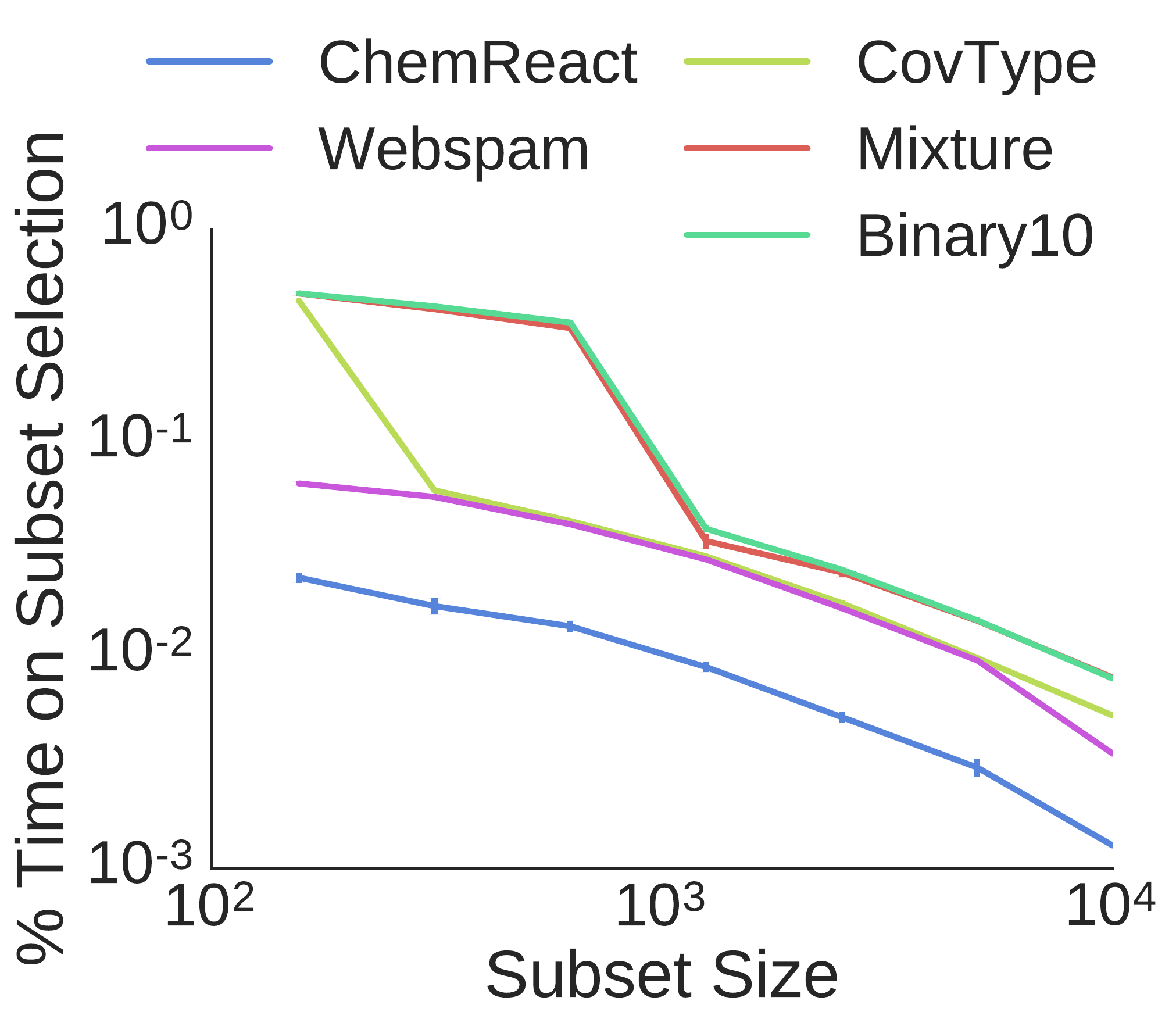}
    \subcaption{}
    \label{fig:relative-time}
\end{subfigure}
\begin{subfigure}[b]{.26\textwidth} %
    \includegraphics[width=\textwidth]{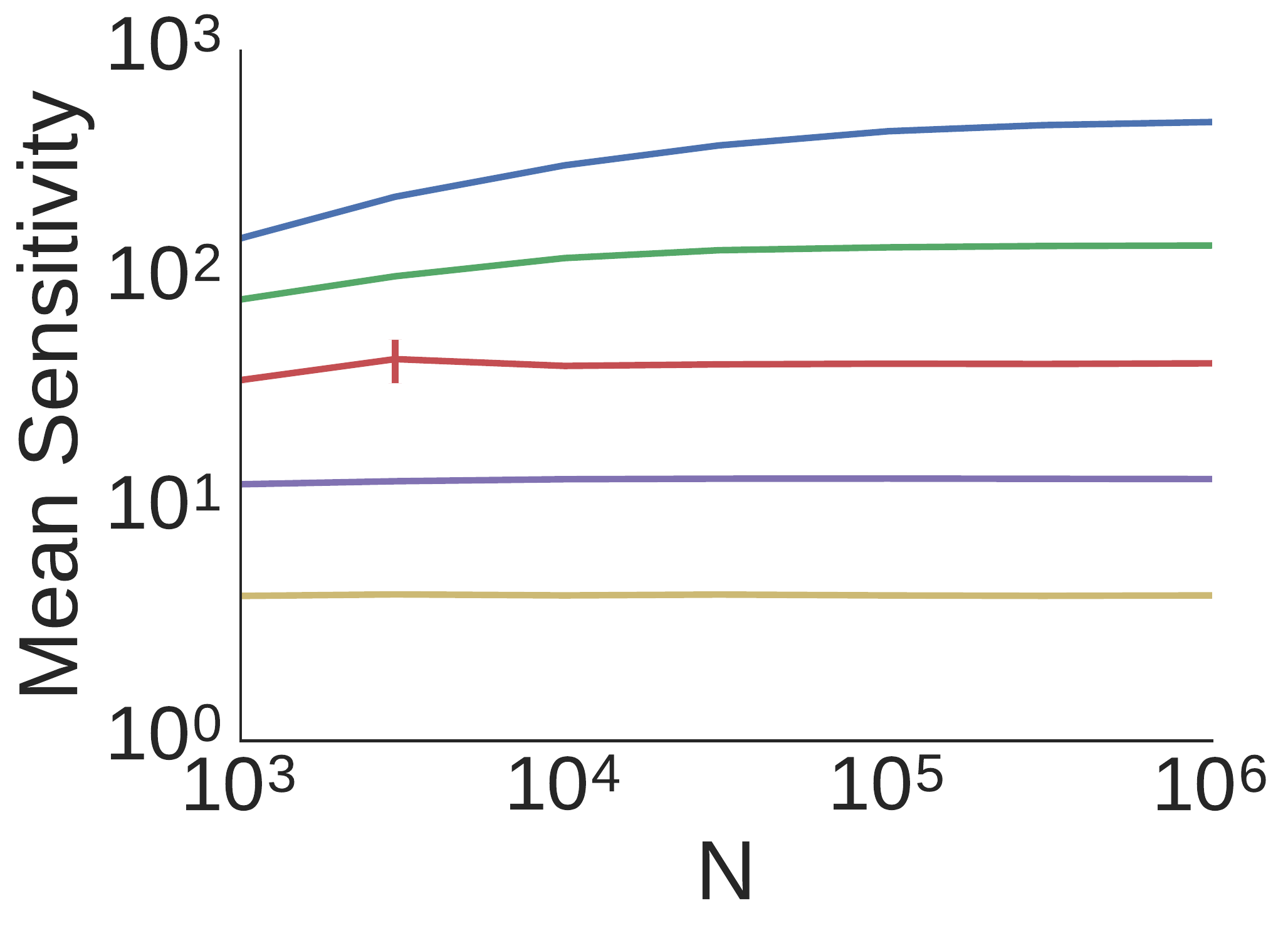} \\
    \includegraphics[width=\textwidth]{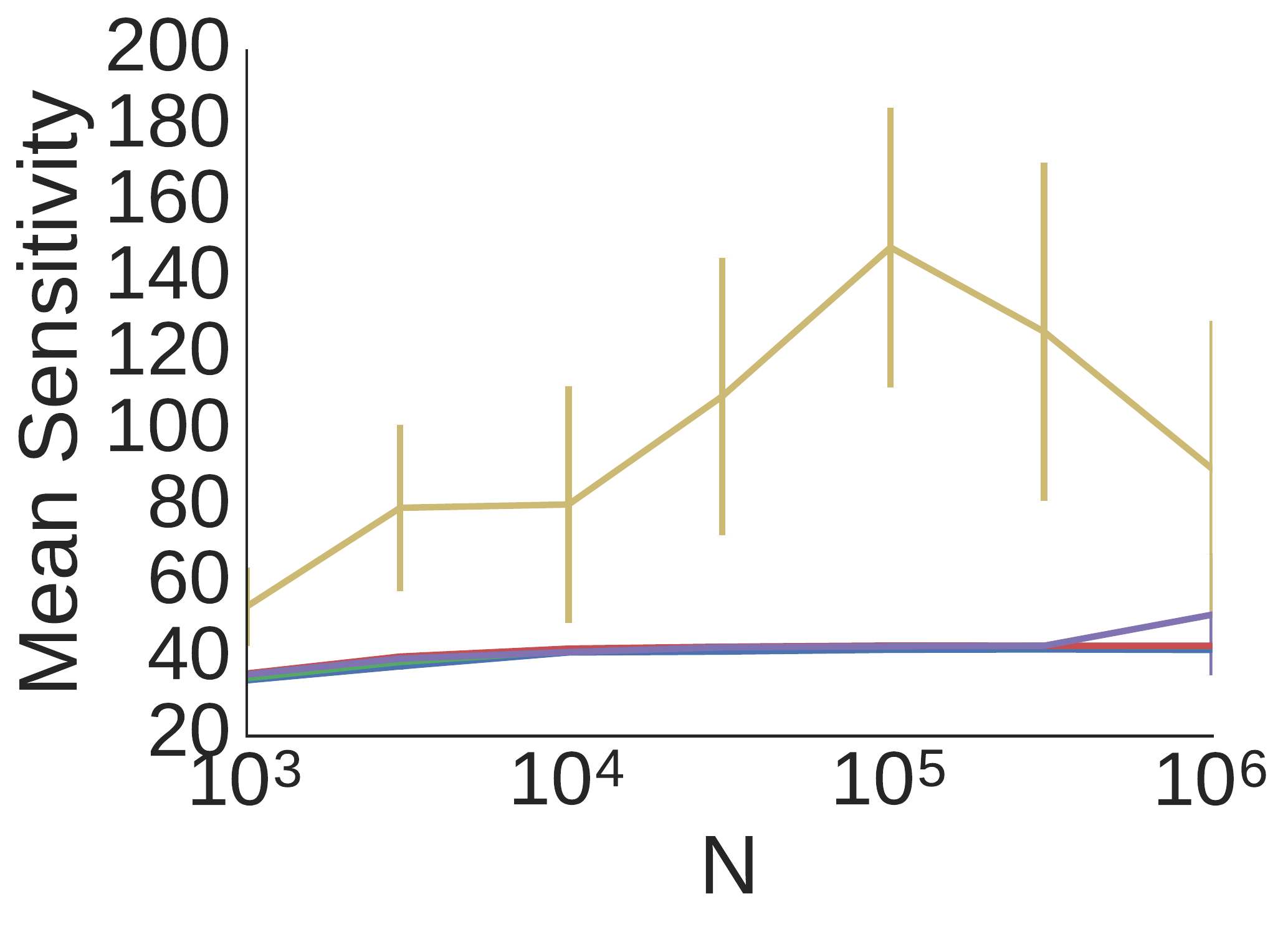}
    \caption{\BinDat}
    \label{fig:synth-cmc-10-sensitivity}
\end{subfigure}
\begin{subfigure}[b]{.29\textwidth} %
    \includegraphics[width=\textwidth]{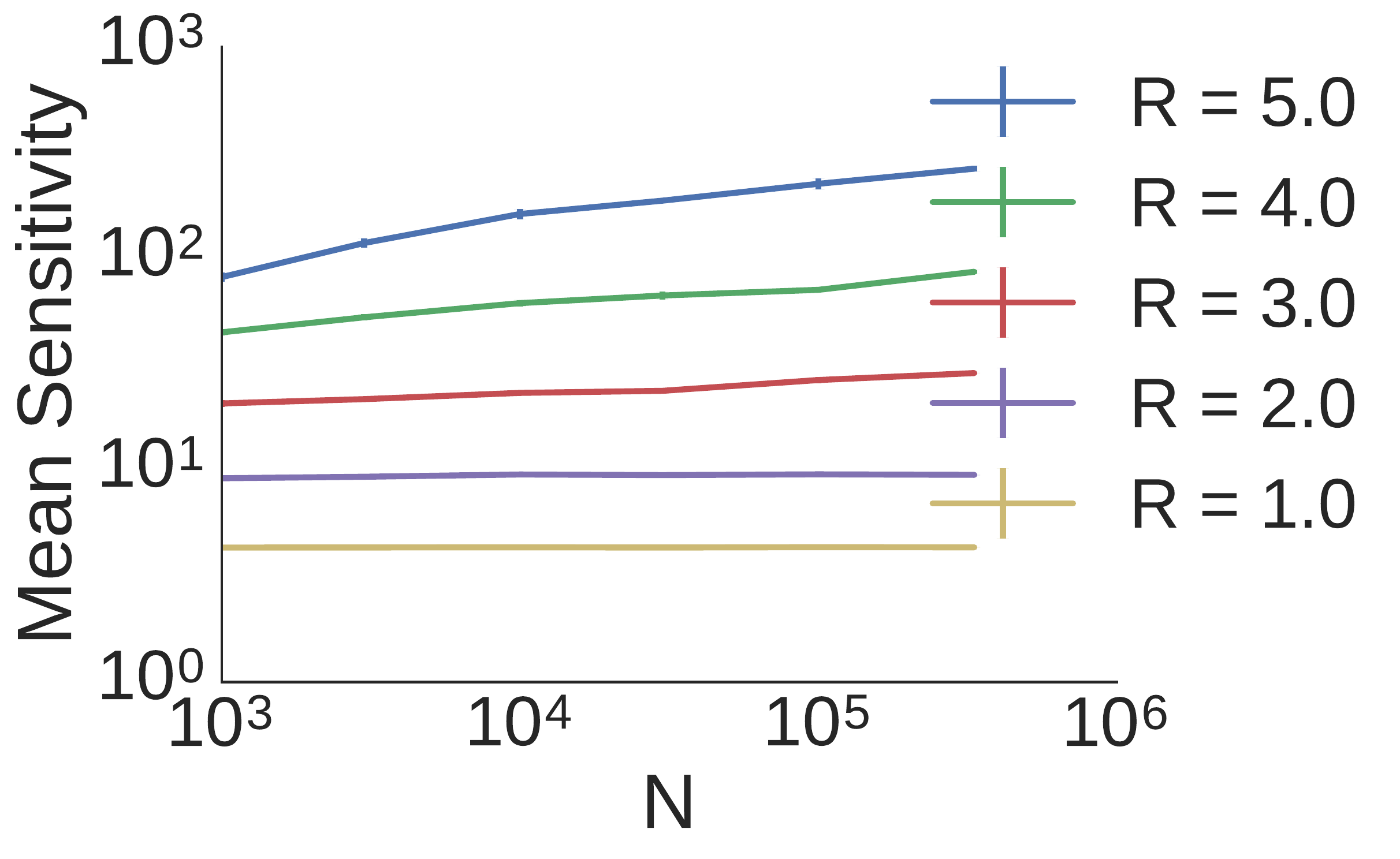} \\
    \includegraphics[width=\textwidth]{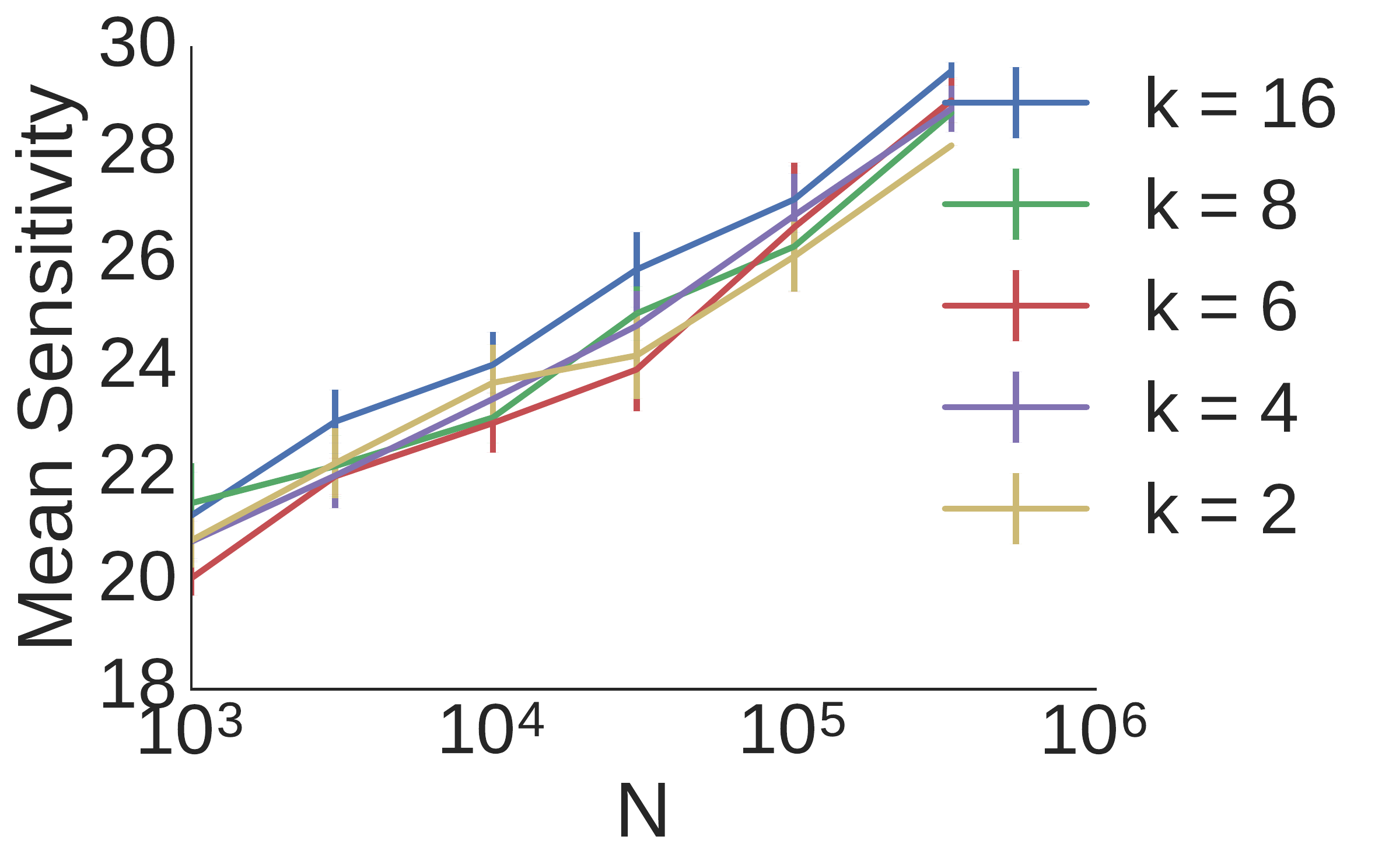}
    \caption{\webspam}
    \label{fig:webspam-sensitivity}
\end{subfigure}
\vspace{.5em}
\caption{\textbf{\opt{arxiv}{(A)}\opt{nips}{(a)}}
Percentage of time spent creating the coreset relative to the total inference time (including 10,000 iterations of MCMC).  
Except for very small coreset sizes, coreset construction is a small fraction of the overall time.
\textbf{\opt{arxiv}{(B,C)}\opt{nips}{(b,c)}}
The mean sensitivities for varying choices of $R$ and $k$.
When $R$ varies $k=6$ and when $k$ varies $R=3$.
The mean sensitivity increases exponentially in $R$, as expected, 
but is robust to the choice of $k$.
}
\label{fig:sensitivities}
\end{center}
\opt{nips}{\vspace{-1em}}
\end{figure}

\opt{arxiv}{
	\newcommand\mainfigscale{.31}
	\newcommand\mainfigscaleup{.343} %
	\newcommand\mainfighspace{0em}
}
\opt{nips}{
	\newcommand\mainfigscale{.25}
	\newcommand\mainfigscaleup{.276} %
	\newcommand\mainfighspace{1em}
}

\begin{figure}[tb]
\begin{center}
\begin{subfigure}[b]{\mainfigscaleup\textwidth}
    \includegraphics[width=\textwidth,clip=true,trim=0 1.4cm 0 0]{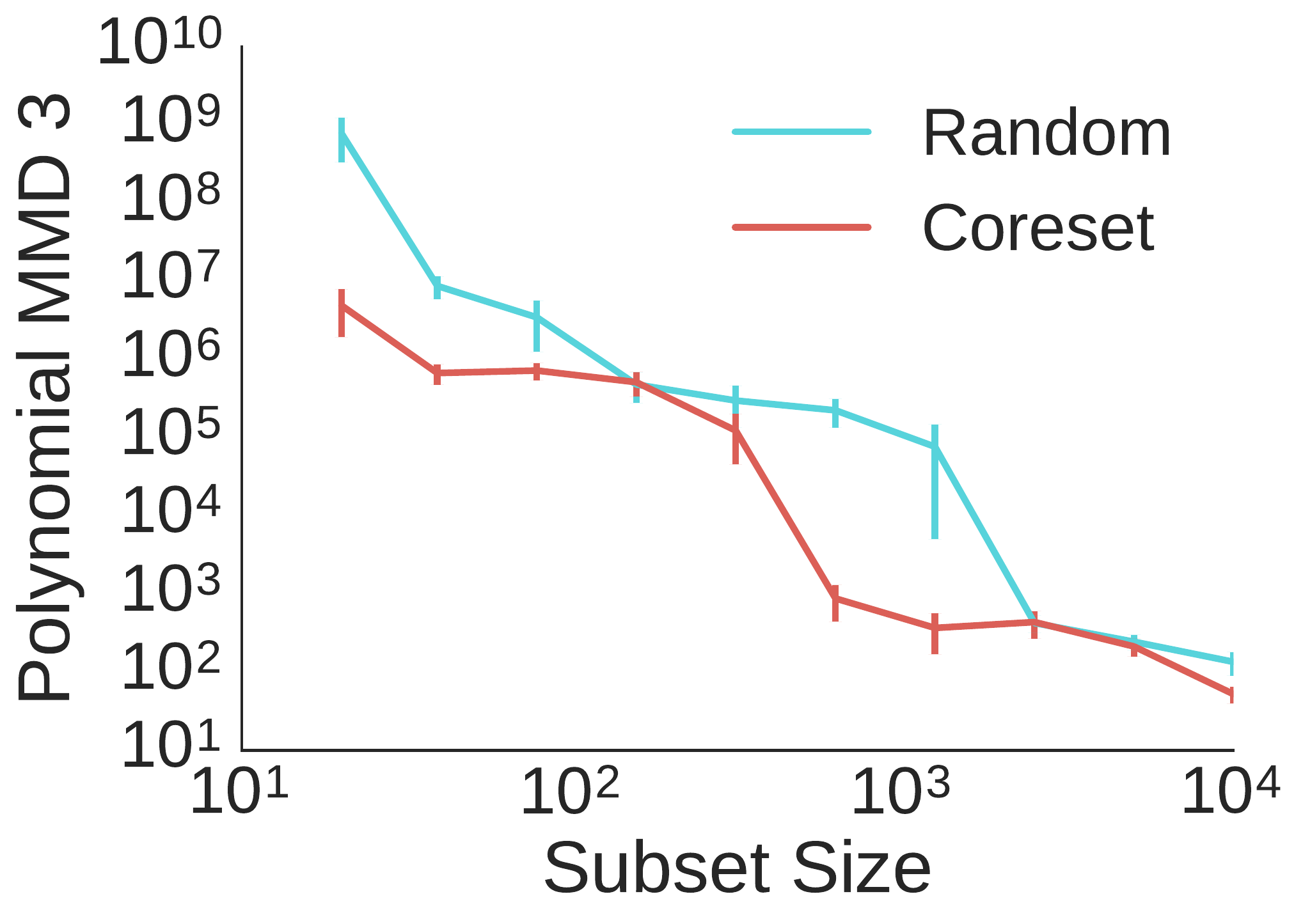} \\
    \includegraphics[width=\textwidth]{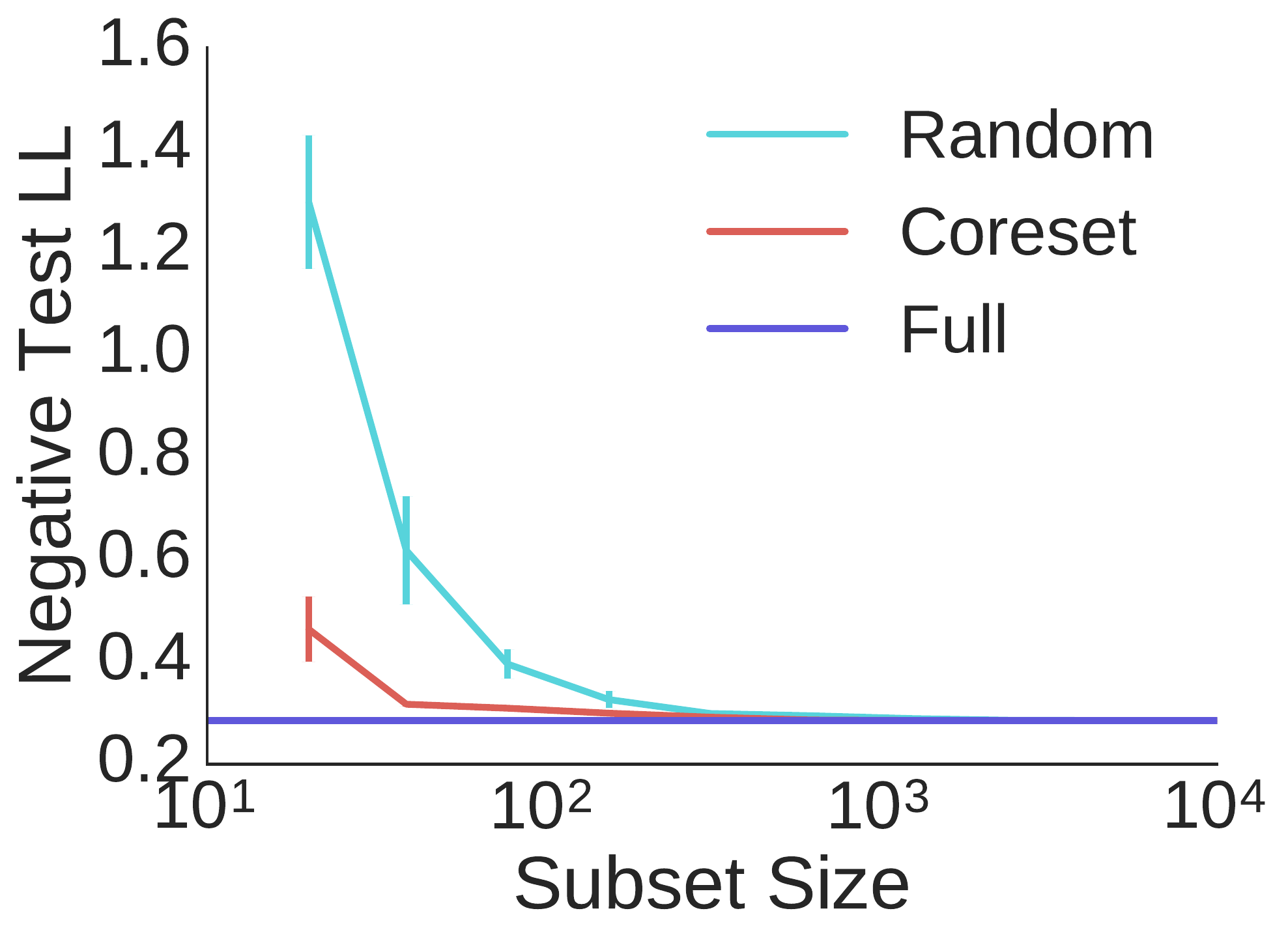}
    \caption{\binDat}
    \label{fig:synth-cmc-5}
\end{subfigure}
\hspace{\mainfighspace}
\begin{subfigure}[b]{\mainfigscale\textwidth}
    \includegraphics[width=\textwidth,clip=true,trim=1.4cm 1.4cm 0 0]{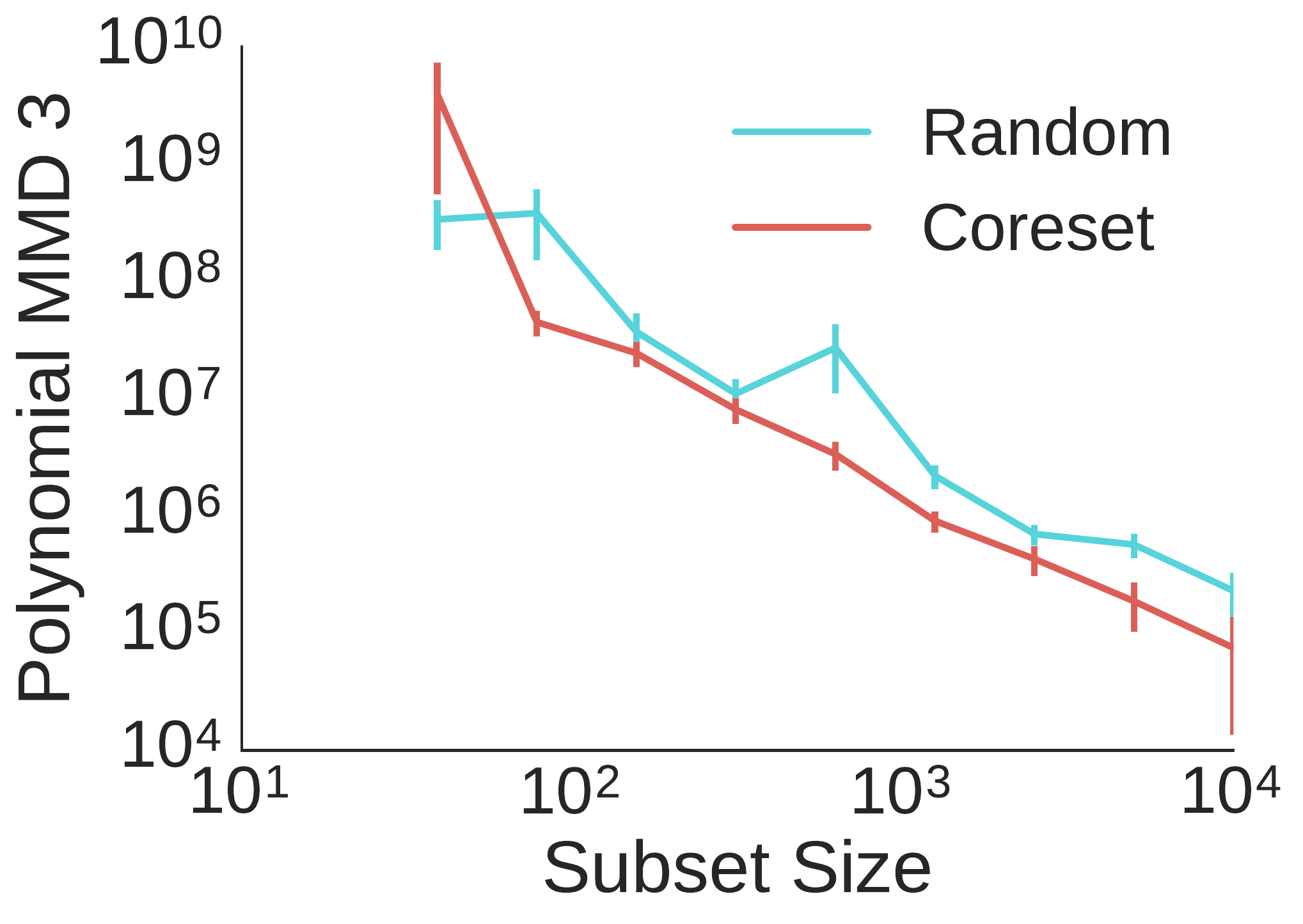} \\
    \includegraphics[width=\textwidth,clip=true,trim=1.4cm 0cm 0 0]{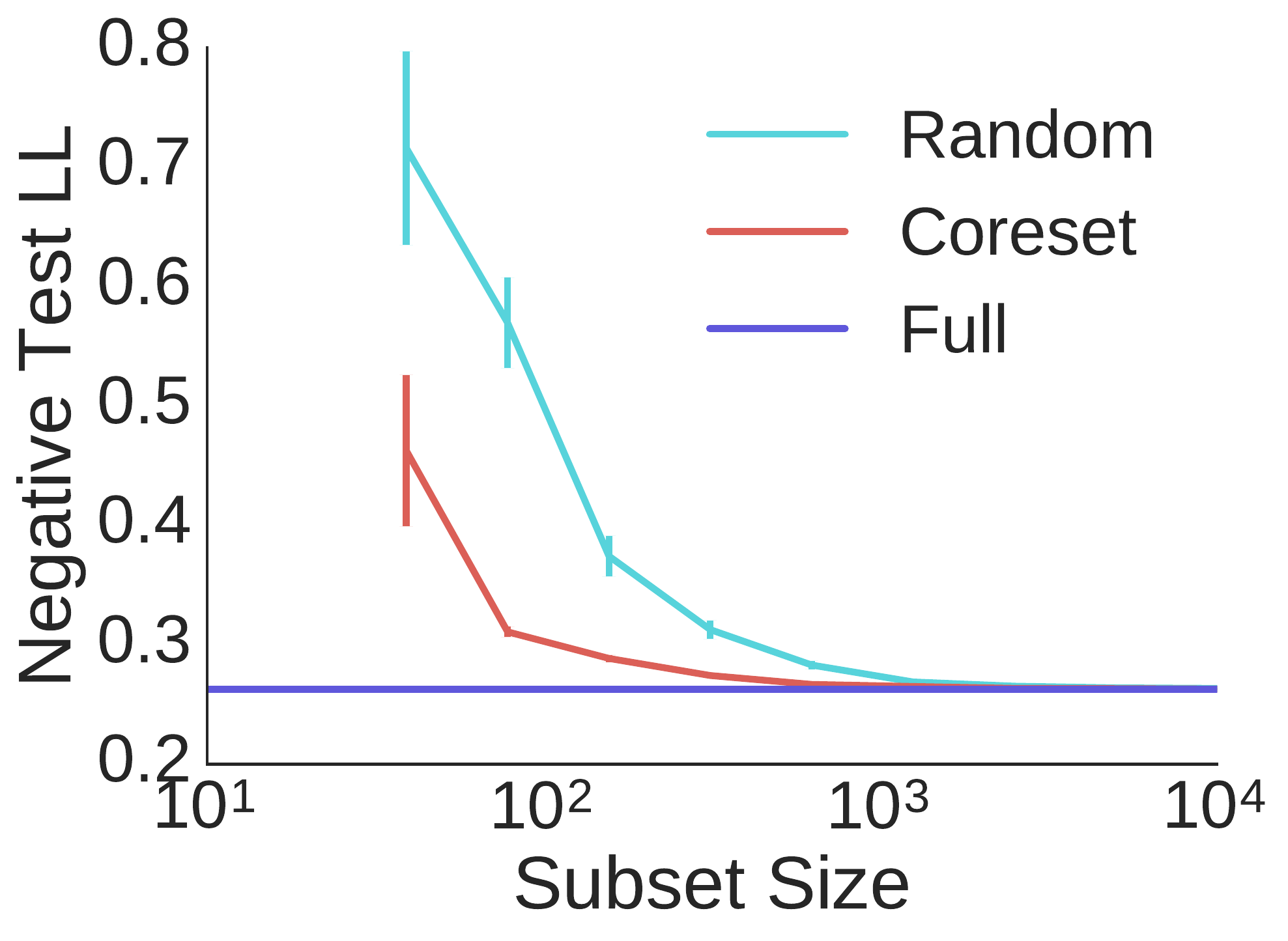}
    \caption{\BinDat}
    \label{fig:synth-cmc-10}
\end{subfigure}
\hspace{\mainfighspace}
\begin{subfigure}[b]{\mainfigscale\textwidth}
    \includegraphics[width=\textwidth,clip=true,trim=1.4cm 1.4cm 0 0]{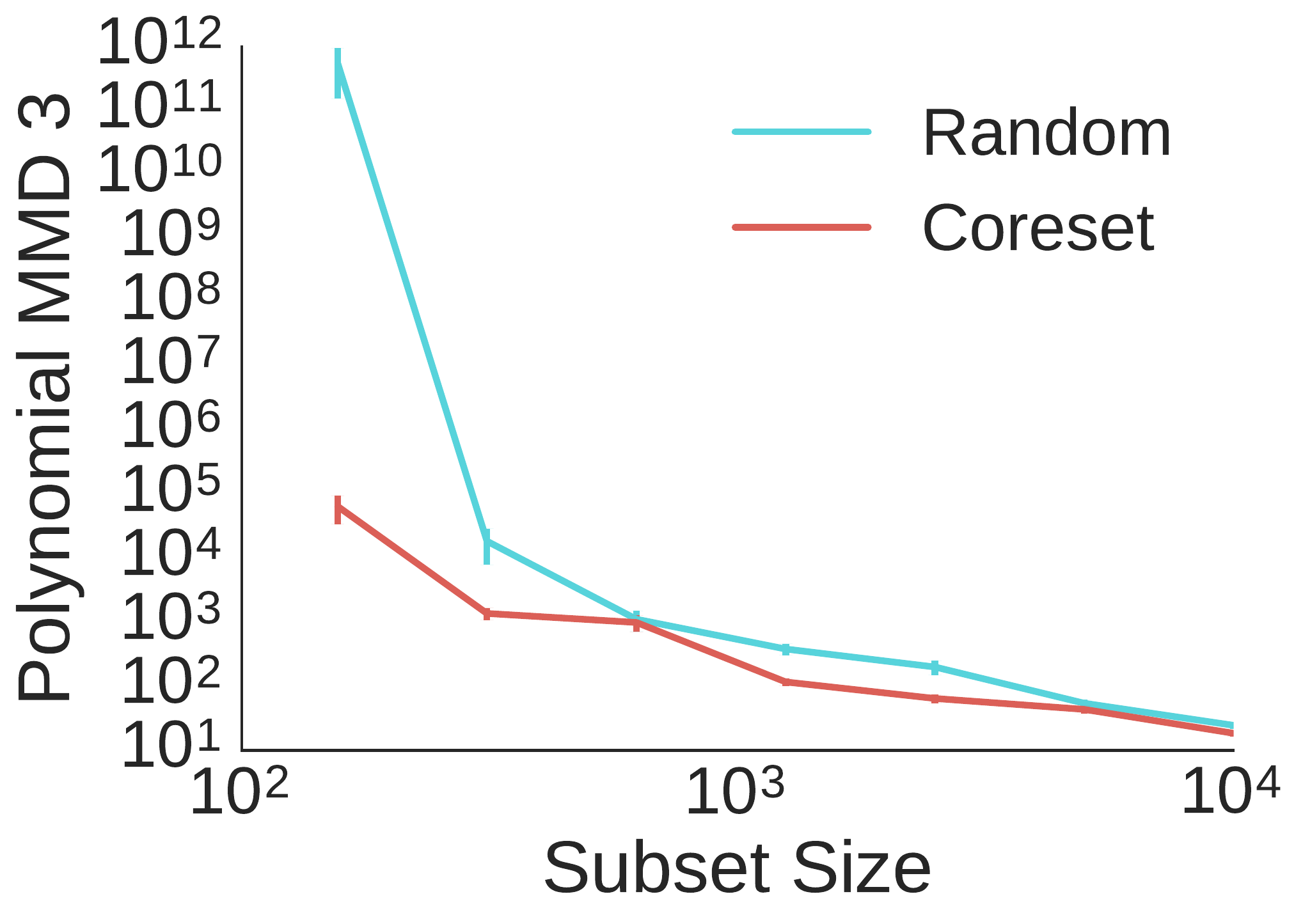} \\
    \includegraphics[width=\textwidth,clip=true,trim=1.4cm 0 0 0]{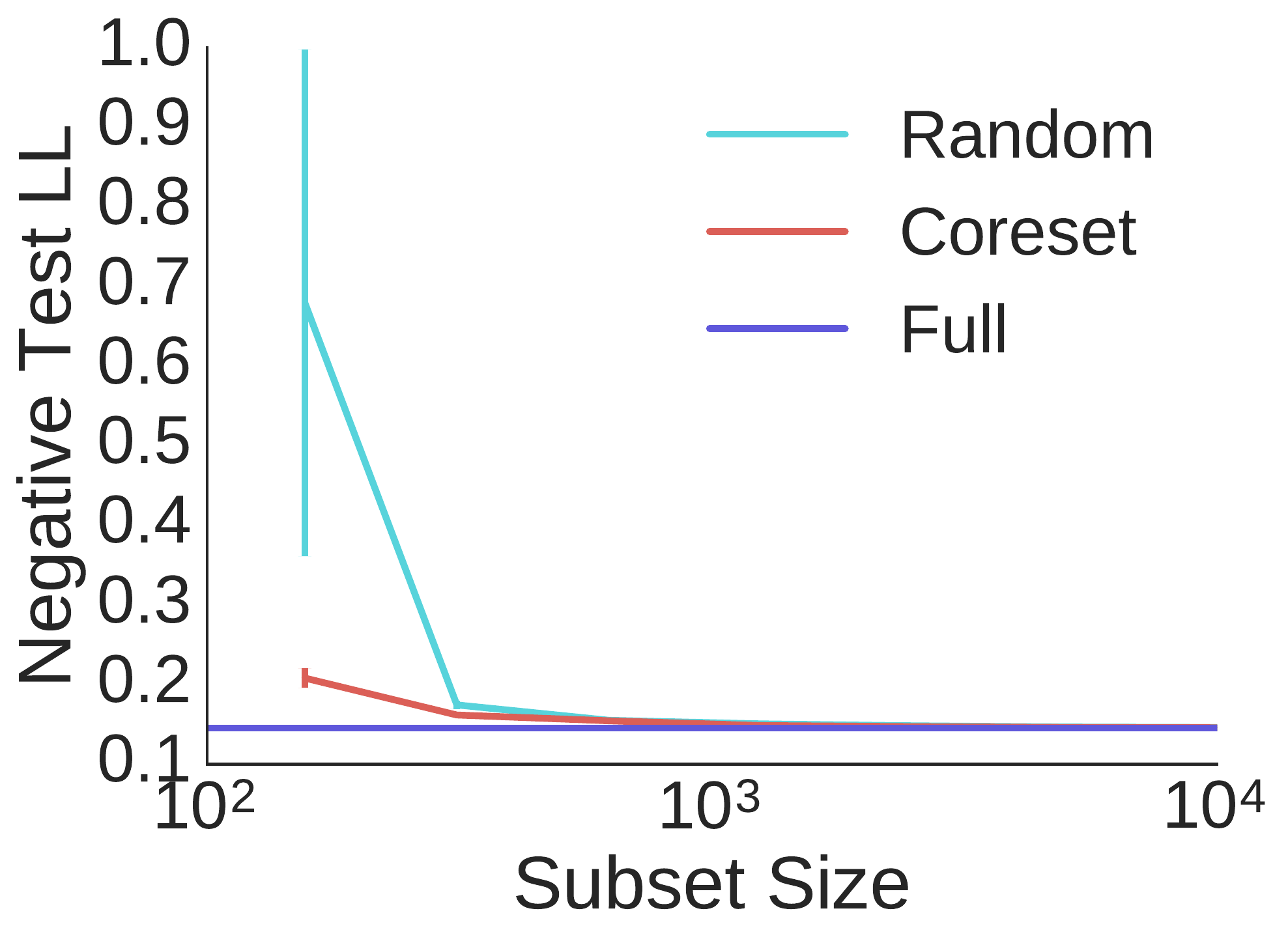}
    \caption{\mixture}
    \label{fig:synth-rev-10}
\end{subfigure} 
\\
\vspace{1em}
\begin{subfigure}[b]{\mainfigscaleup\textwidth}
    \includegraphics[width=\textwidth,clip=true,trim=0 1.4cm 0 0]{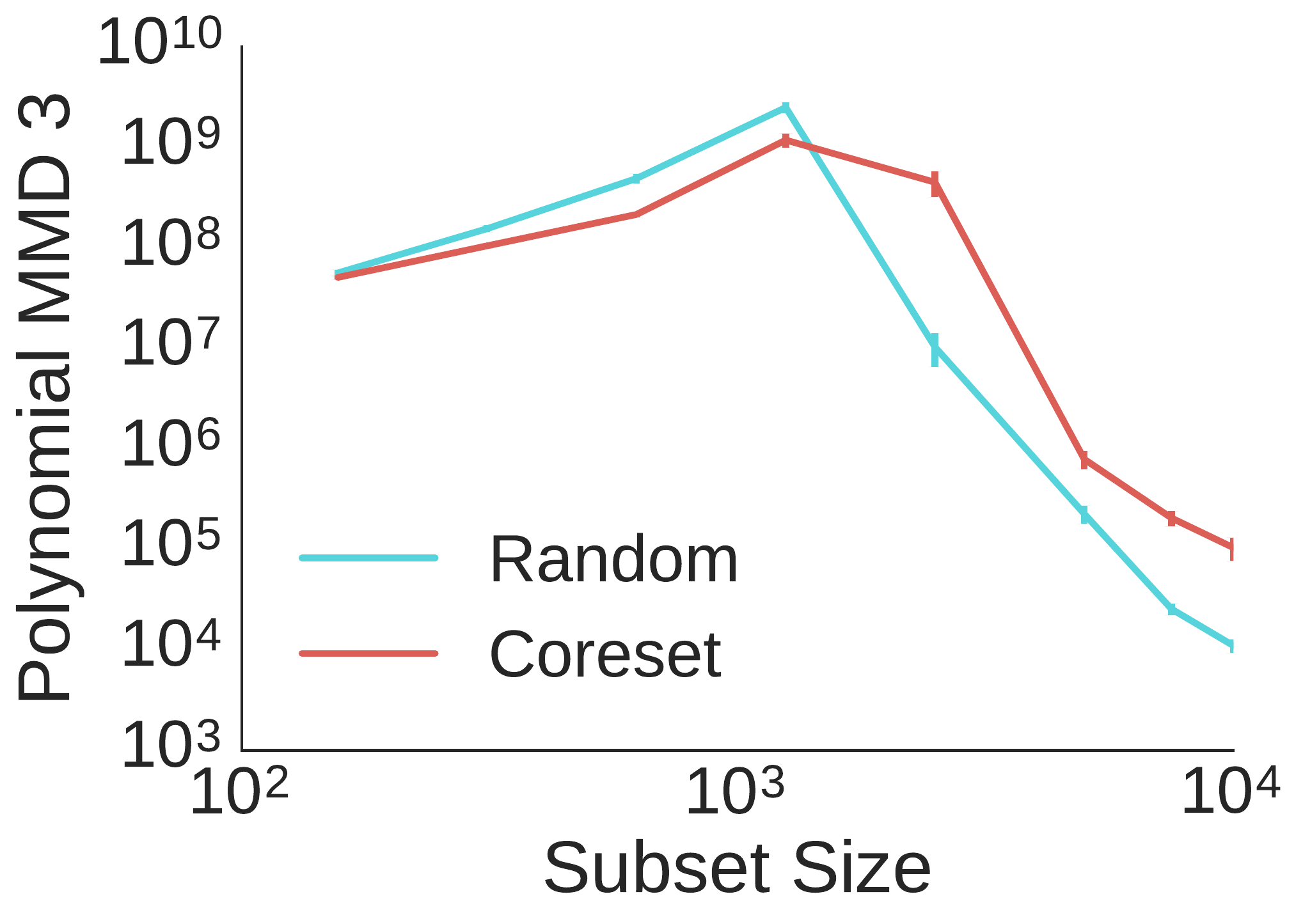} \\
    \includegraphics[width=\textwidth]{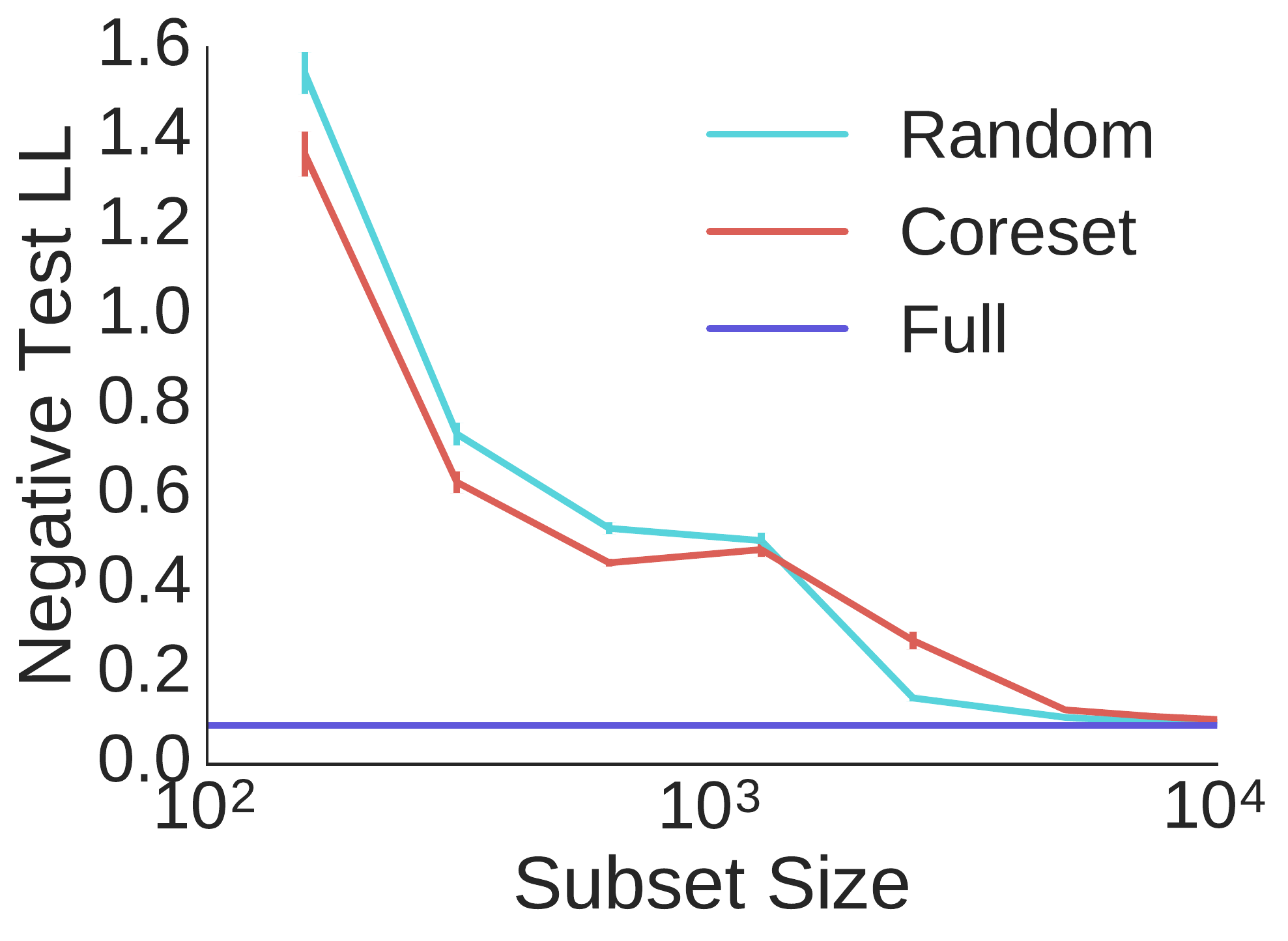}
    \caption{\chemre}
    \label{fig:chemreact}
\end{subfigure}
\hspace{\mainfighspace}
\begin{subfigure}[b]{\mainfigscale\textwidth}
    \includegraphics[width=\textwidth,clip=true,trim=1.4cm 1.4cm 0 0]{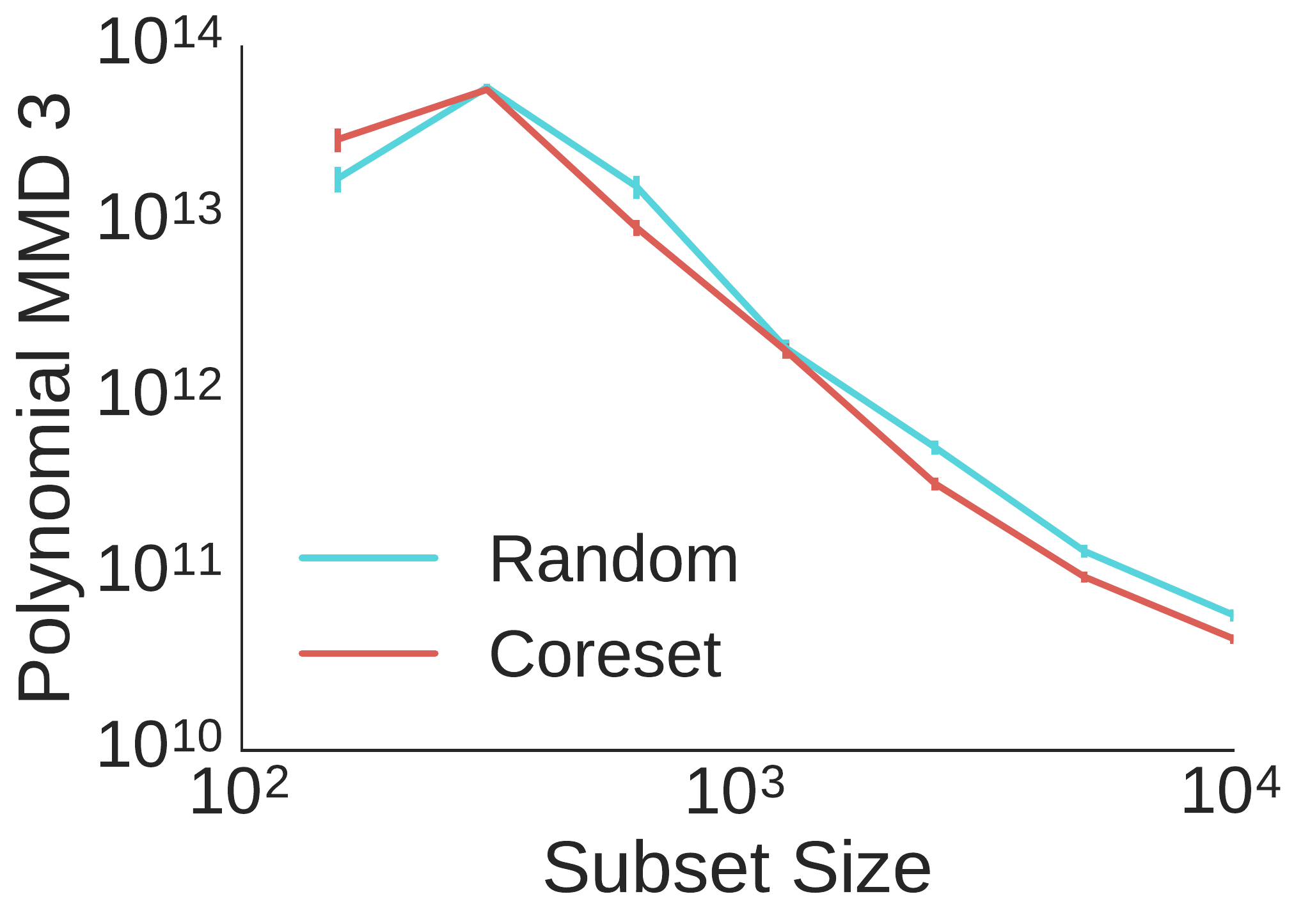} 
    \includegraphics[width=\textwidth,clip=true,trim=1.4cm 0cm 0 0]{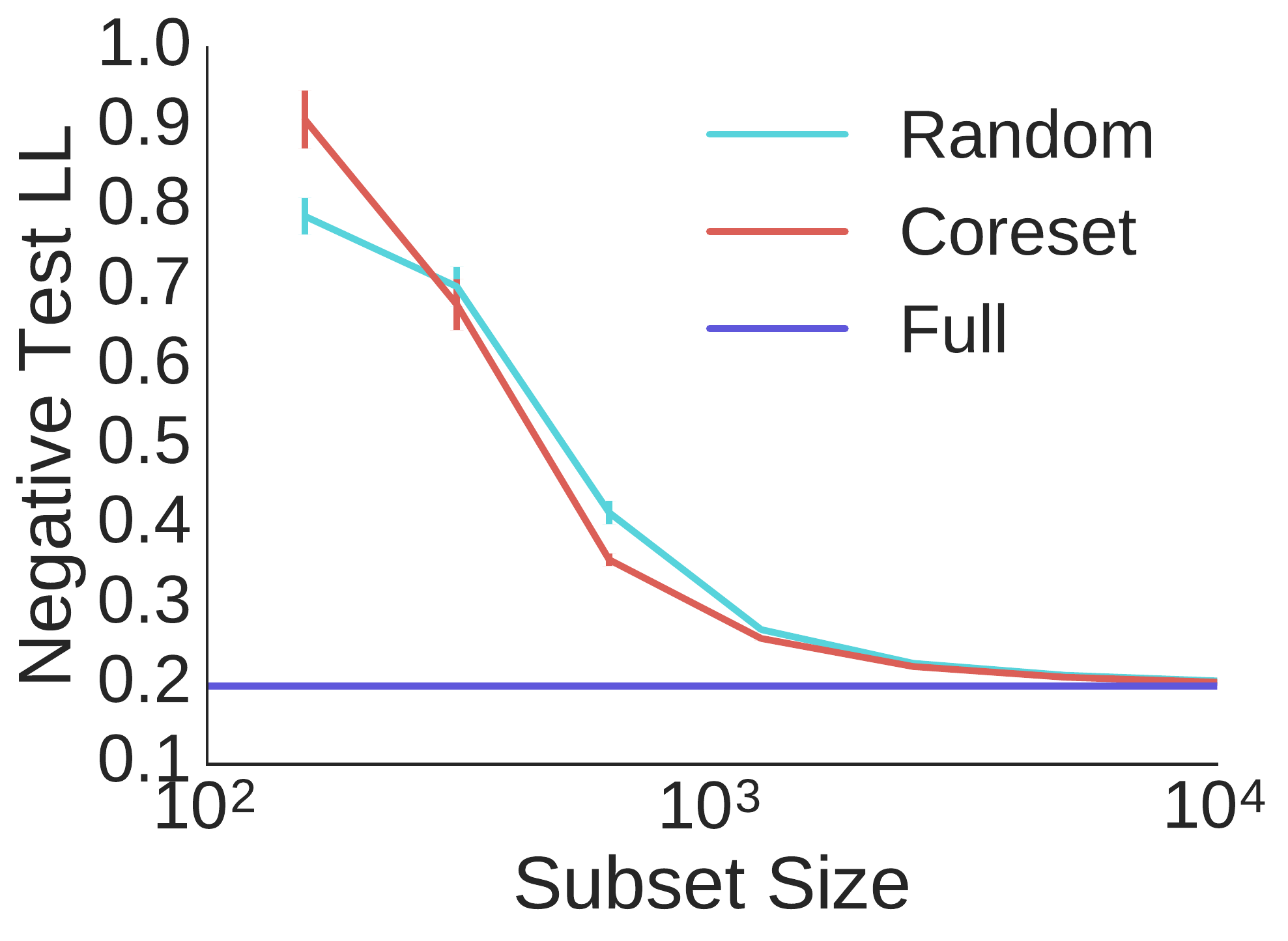}
    \caption{\webspam}
    \label{fig:webspam}
\end{subfigure}
\hspace{\mainfighspace}
\begin{subfigure}[b]{\mainfigscale\textwidth}
    \includegraphics[width=\textwidth,clip=true,trim=1.4cm 1.4cm 0 0]{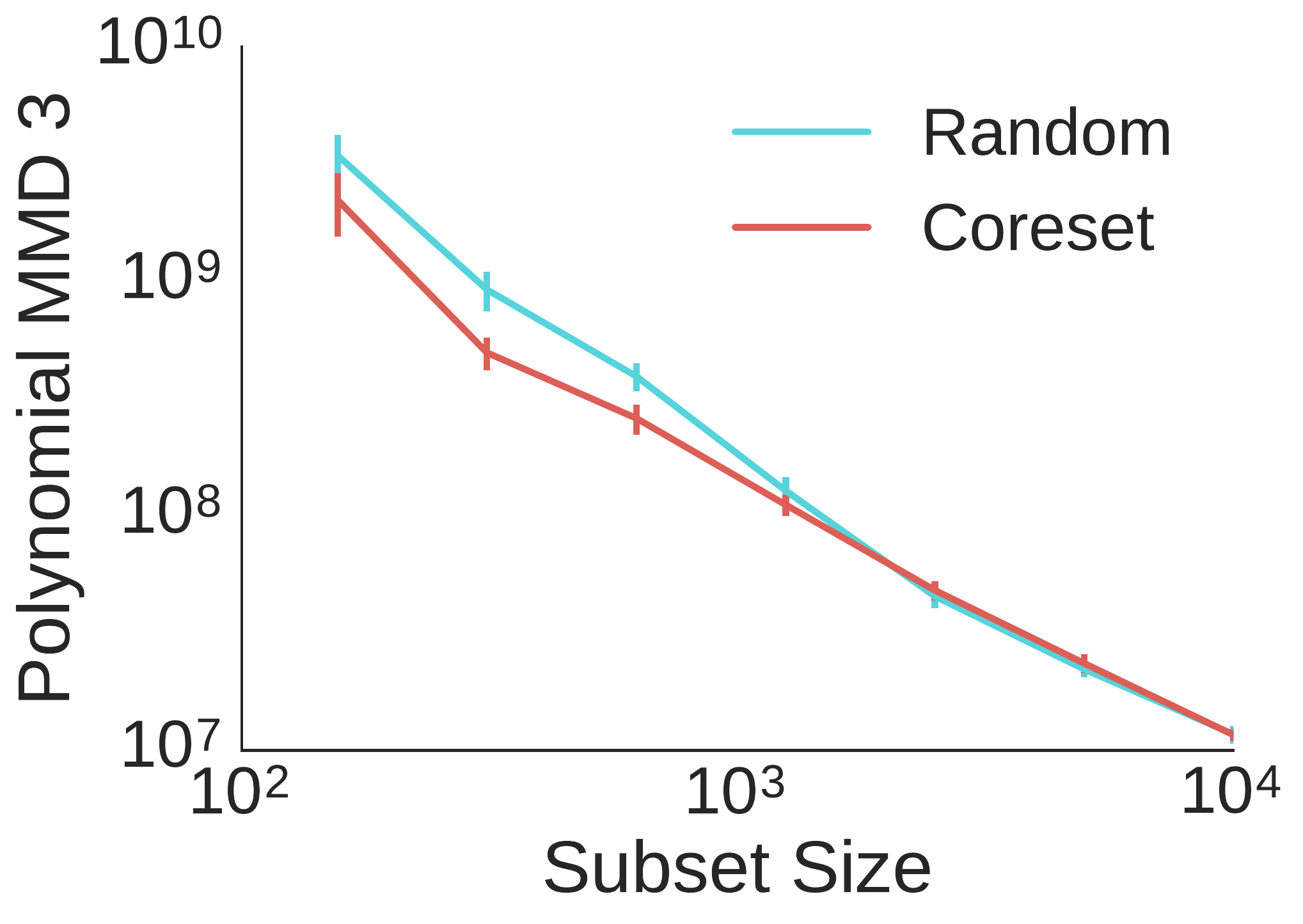} 
    \includegraphics[width=\textwidth,clip=true,trim=1.4cm 0cm 0 0]{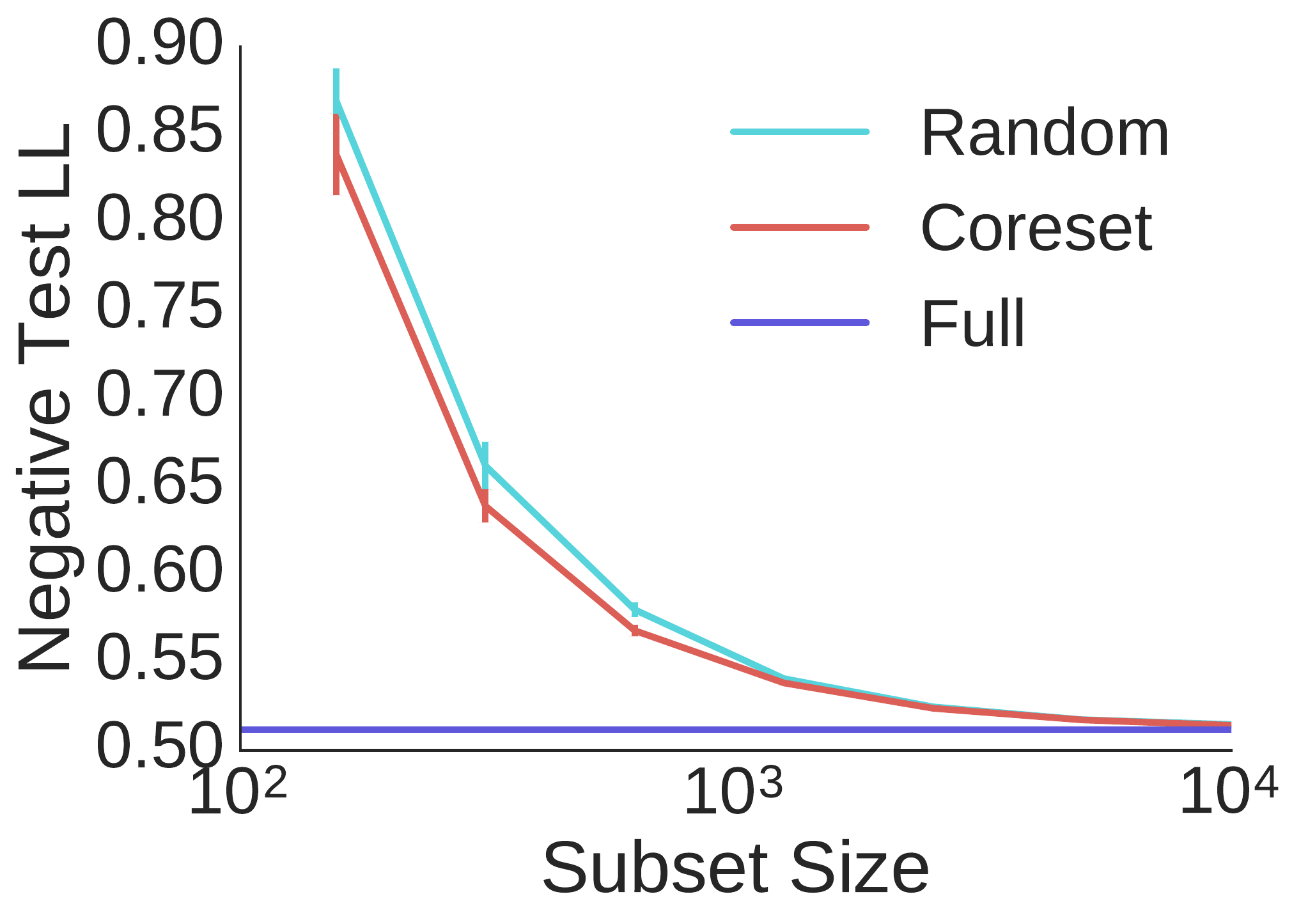}
    \caption{\covtype}
    \label{fig:covtype}
\end{subfigure}
\opt{nips}{\vspace{.5em}}
\caption{Polynomial MMD and negative test log-likelihood of random sampling and the 
logistic regression coreset algorithm for synthetic and real data with varying 
subset sizes (lower is better for all plots). 
For the synthetic data, $N=10^{6}$ total data points were used and 
$10^{3}$ additional data points were generated for testing. 
For the real data, 2,500 (resp.~50,000 and 29,000) data points of the \chemre 
(resp.~\webspam and \covtype) dataset were held out for testing.
One standard deviation error bars were obtained by repeating each experiment 20 times.
}
\label{fig:main-results}
\end{center}
\opt{nips}{\vspace{-1em}}
\end{figure}
We evaluated the performance of the logistic regression coreset algorithm on a number 
of synthetic and real-world datasets. 
We used a maximum dataset size of 1 million examples
because we wanted to be able to 
calculate the true posterior, which would be infeasible for extremely large datasets.

\textbf{Synthetic Data.} 
We generated synthetic binary data according to the model
$X_{nd} \distind \distBern(p_{d}), d=1,\dots,D$ and $Y_{n} \distind p_{logistic}(\cdot \given X_{n}, \theta)$. 
The idea is to simulate data in which there are a small number of rarely occurring but highly
predictive features, which is a common real-world phenomenon.
We thus took $p = (1,  .2,  .3, .5, .01,  .1,  .2, .007, .005, .001)$ and
$\theta = (-3, 1.2, -.5, .8,  3, -1.,-.7,  4,  3.5,  4.5)$ for the $D=10$ experiments (\BinDat) and 
the first 5 components of $p$ and $\theta$ for the $D=5$ experiments (\binDat). 
The generative model is the same one used by \citet{Scott:2013} and the first 5 components of
$p$ and $\theta$ correspond to those used in the Scott et al.~experiments (given in \citep[Table 1b]{Scott:2013}). 
We generated a synthetic mixture dataset with continuous covariates (\mixture)  
using a model similar to that of \citet{Han:2016}: $Y_{n} \distiid \distBern(1/2)$
and $X_{n} \distind \distNorm(\mu_{Y_{n}}, I)$, where $\mu_{-1} = (0,0,0,0,0,1,1,1,1,1)$
and $\mu_{1} = (1,1,1,1,1,0,0,0,0,0)$.

\textbf{Real-world Data.}
The \chemre dataset consists of $N=$ 26,733 chemicals, each with $D=100$ properties. 
The goal is to predict whether each chemical is reactive. 
The \webspam corpus consists of $N=$ 350,000 web pages, approximately 60\% of which are spam. 
The covariates consist of the $D=127$ features that each appear in at least 25 documents. 
The cover type (\covtype) dataset consists of $N=$ 581,012 cartographic observations with $D=54$ features.
The task is to predict the type of trees that are present at each observation location.

\subsection{Scaling Properties of the Coreset Construction Algorithm}

\textbf{Constructing Coresets.} 
In order for coresets to be a worthwhile preprocessing step, it is critical that the 
time required to construct the coreset is small relative to the time needed to complete the
inference procedure. 
We implemented the logistic regression coreset algorithm in Python.\footnote{More details on 
our implementation are provided in the \suppmat. 
Code to recreate all of our experiments is available at 
\url{https://bitbucket.org/jhhuggins/lrcoresets}.}
In \cref{fig:relative-time}, we plot the relative time to construct the coreset for each type of dataset ($k=6$)
versus the total inference time, including 10,000 iterations of the MCMC procedure described in \cref{sec:posterior-approx-quality}.
Except for very small coreset sizes, the time to run MCMC dominates. 

\textbf{Sensitivity.} 
An important question is how the mean sensitivity $\bbm_{N}$ scales with $N$, as
it determines how the size of the coreset scales with the data. 
Furthermore, ensuring that mean sensitivity is robust to the number of clusters $k$ 
is critical since needing to adjust the algorithm hyperparameters for each dataset 
could lead to an unacceptable increase in computational burden. 
We also seek to understand how the radius $R$ affects the mean sensitivity. 
\cref{fig:synth-cmc-10-sensitivity,fig:webspam-sensitivity} show the results of 
our scaling experiments on the \BinDat and \webspam data. 
The mean sensitivity is essentially constant across a range of dataset sizes.
For both datasets the mean sensitivity is robust to the choice of $k$ and
scales exponentially in $R$, as we would expect from \cref{lem:sensitivity-upper-bound}.
\opt{nips}{\vspace{-.47em}}

\subsection{Posterior Approximation Quality}
\label{sec:posterior-approx-quality}

Since the ultimate goal is to use coresets for  Bayesian inference,
the key empirical question is how well a posterior
formed using a coreset approximates the true posterior distribution. 
We compared the coreset algorithm to random subsampling of data points,
since that is the approach used in many existing scalable versions of variational 
inference and MCMC~\opt{arxiv}{\citep{Hoffman:2013,Bardenet:2014,Korattikara:2014,Bardenet:2015}}\opt{nips}{\citep{Hoffman:2013,Bardenet:2015}}.
Indeed, coreset-based importance sampling could be used as 
a drop-in replacement for the random subsampling used by these methods,
though we leave the investigation of this idea for future work. 

\textbf{Experimental Setup.} 
We used adaptive Metropolis-adjusted Langevin algorithm (MALA)~\citep{Roberts:1996,Haario:2001}
for posterior inference.
For each dataset, we ran the coreset and random subsampling algorithms 20 times for each choice of 
subsample size $M$. 
We ran adaptive MALA for 100,000 iterations on the full dataset and each subsampled dataset.
The subsampled datasets were fixed for the entirety of each run, in contrast to subsampling 
algorithms that resample the data at each iteration. 
For the synthetic datasets, which are lower dimensional, we used $k = 4$ while
for the real-world datasets, which are higher dimensional, we used $k = 6$.
We used a heuristic to choose $R$ as large as was feasible while still obtaining 
moderate total sensitivity bounds. 
For a clustering $\mcQ$ of data $\mcD$, let $\mcI \defined N^{-1}\sum_{i=1}^{k}\sum_{Z \in G_{i}}\|Z - Q_{i}\|^{2}$
be the normalized $k$-means score. 
We chose $R = a/\sqrt{\mcI}$, where $a$ is a small constant.
The idea is that, for $i \in [k]$ and $Z_{n} \in G_{i}$, we want $R\|\bZ_{G,i}^{(-n)} - Z_{n}\|_{2} \approx a$ 
on average, so the term $\exp\{-R\|\bZ_{G,i}^{(-n)} - Z_{n}\|_{2}\}$ in \cref{eq:sensitivity-upper-bound}
is not too small and hence $\sigma_{n}(\ball_{R})$ is not too large. 
Our experiments used $a = 3$. 
We obtained similar results for $4 \le k \le 8$ and $2.5 \le a \le 3.5$,
indicating that the logistic regression coreset algorithm has some robustness to the choice 
of these hyperparameters.
We used negative test log-likelihood and maximum mean discrepancy (MMD) with a
3rd degree polynomial kernel as comparison metrics (so smaller is better).

\textbf{Synthetic Data Results.} 
Figures \ref{fig:synth-cmc-5}-\ref{fig:synth-rev-10} show the results for synthetic data.
In terms of test log-likelihood, coresets did as well as or outperformed random subsampling.
In terms of MMD, the coreset posterior approximation typically outperformed random 
subsampling by 1-2 orders of magnitude and never did worse.
These results suggest much can be gained by using coresets, with comparable 
performance to random subsampling in the worst case. 

\textbf{Real-world Data Results.} 
Figures \ref{fig:chemreact}-\ref{fig:covtype} show the results for real data.
Using coresets led to better performance on \chemre for small subset sizes.
Because the dataset was fairly small and random subsampling was done without replacement, 
coresets were worse for larger subset sizes.
Coreset and random subsampling performance was approximately the same for \webspam.
On \webspam and \covtype, coresets either outperformed or did as well as random subsampling in terms
MMD and test log-likelihood on almost all subset sizes.
The only exception was that random subsampling was superior on \webspam for the smallest subset set.
We suspect this is due to the variance introduced by the importance sampling procedure used to 
generate the coreset. 

For both the synthetic and real-world data, in many cases we are able to obtain 
a high-quality logistic regression posterior approximation using a coreset that
is many orders of magnitude smaller than the full dataset -- sometimes just a few hundred
data points. 
Using such a small coreset represents a substantial reduction in the memory and computational
requirements of the Bayesian inference algorithm that uses the coreset for posterior inference. 
We expect that the use of coresets could lead similar gains for other Bayesian models.
Designing coreset algorithms for other widely-used models is an exciting direction for future research.  

%

%% file: appendix.tex
\appendix

\numberwithin{equation}{section}

\section{Marginal Likelihood Approximation}

\begin{proofof}{\cref{prop:marginal-likelihood-approx}}
By the assumption that $\mcL$ and $\tilde\mcL$ are non-positive, the multiplicative error
assumption, and Jensen's inequality,
\[
\tilde\mcE
&= \int e^{\tilde\mcL(\theta)}\pi_{0}(\theta)\,\dee \theta 
\ge \int e^{(1+\varepsilon)\mcL(\theta)}\pi_{0}(\theta)\,\dee \theta 
\ge \left(\int e^{\mcL(\theta)}\pi_{0}(\theta)\,\dee \theta\right)^{1+\varepsilon} 
= \mcE^{1+\varepsilon}
\]
and
\[
\tilde\mcE
&= \int e^{\tilde\mcL(\theta)}\pi_{0}(\theta)\,\dee \theta 
\le \int e^{(1-\varepsilon)\mcL(\theta)}\pi_{0}(\theta)\,\dee \theta 
\le \left(\int e^{\mcL(\theta)}\pi_{0}(\theta)\,\dee \theta\right)^{1-\varepsilon} 
= \mcE^{1-\varepsilon}.
\]
\end{proofof}

\section{Main Results}

In order to construct coresets for logistic regression, we will use the framework
developed by \citet{Feldman:2011a} and improved upon by \citet{Braverman:2016}. 
For $n \in [N] \defined \theset{1,\dots,N}$, let $f_{n} : \mcS \to \reals_{+}$
be a non-negative function from some set $\mcS$ and let 
$\bbf = \frac{1}{N} \sum_{n=1}^{N} f_{n}$ be the average of the functions. 
Define the \emph{sensitivity} of $n \in [N]$ with respect to $\mcS$ by
\[
\sigma_{n}(\mcS) \defined \sup_{s \in S} \frac{f_{n}(s)}{\bbf(s)},
\]
and note that $\sigma_{n}(\mcS) \le N$. 
Also, for the set $\mcF \defined \{ f_{n} \given n \in [N] \}$, define the dimension
$\dim(\mcF)$ of $\mcF$ to be the minimum integer $d$ such that 
\[
\forall F \subseteq \mcF,\;|\{F \cap R \given R \in \ranges(\mcF)\}| \le (|F|+1)^{d},
\]
where $\ranges(\mcF) \defined \{\range(s, a) | s \in \mcS, a \ge 0 \}$ and 
$\range(s,a) \defined \{ f \in \mcF \given f(s) \le a \}$. 

We make use of the following improved version of \citet[Theorems 4.1 and 4.4]{Feldman:2011a}.
\bnthm[{\citet{Braverman:2016}}] \label{thm:making-coresets}
Fix $\varepsilon > 0$. 
For $n \in [N]$, let $m_{n} \in \reals_{+}$ be chosen such that 
\[
m_{n} \ge \sigma_{n}(\mcS)
\]
and let $\bbm_N \defined \frac{1}{N}\sum_{n=1}^{N} m_{n}$. 
There is a universal constant $c$ such that if $\mcC$ is a 
sample from $\mcF$ of size 
\[
|\mcC| \ge \frac{c\,\bbm_N}{\varepsilon^{2}}(\dim(\mcF)\log\bbm_{N} + \ln(1/\delta)),
\]
such that the probability that each element of $\mcC$ is selected independently
from $\mcF$ with probability $\frac{m_{n}}{N\bbm_{N}}$ that $f_{n}$ is chosen, 
then with probability at least $1 - \delta$, for all $s \in \mcS$,
\[
\left|\bbf(s) - \frac{\bbm_{N}}{|\mcC|} {\textstyle\sum_{f \in \mcC}} \frac{f(s)}{m_{n}}\right| \le \varepsilon \bbf(s).  
\]
\enthm

The set $\mcC$ in the theorem is called a \emph{coreset}. 
In our application to logistic regression, $\mcS = \Theta$ and
$f_{n}(\theta) = -\ln p(Y_{n} \given X_{n}, \theta)$.
The key is to determine $\dim(\mcF)$ and to construct the values $m_{n}$ efficiently.
Furthermore, it is necessary for $\bbm_N = o(\sqrt{N})$ at a minimum and preferable
for $\bbm_N = O(1)$. 

Letting $Z_{n} = Y_{n}X_{n}$ and $\phi(s) = \ln(1 + \exp(-s))$, we can rewrite 
$f_{n}(\theta) = \phi(Z_{n} \cdot \theta)$. 
Hence, the goal is to find an upper bound 
\[
m_{n} 
\ge \sigma_{n}(\Theta) 
= \sup_{\theta \in \Theta} \frac{N\,\phi(Z_{n} \cdot \theta)}{\sum_{n'=1}^{N} \phi(Z_{n'} \cdot \theta)}.
\]

To obtain an upper bound on the sensitivity, we will take 
$\Theta = \ball_{R}$ for some $R > 0$.

\bnlem \label{lem:phi-ratio-bound}
For all $a,b \in \reals$, $\phi(a)/\phi(b) \le e^{|a-b|}$.
\enlem
\bprf
The lemma is trivial when $a = b$. 
Let $\Delta = b - a \ne 0$ and $\rho(a) = \phi(a)/\phi(a+\Delta)$. 
We have
\[
\rho'(a) = \frac{(1+e^{a})\log(1+e^{-a})-(1+e^{a+\Delta})\log(1+e^{-a-\Delta})}{(1+e^{a})(1+e^{a+\Delta})\log^{2}(1+e^{-a-\Delta})}.
\]
Examining the previous display we see that $\sgn(\rho'(a)) = \sgn(\Delta)$.
Hence if $\Delta > 0$, 
\[
\sup_{a} \frac{\phi(a)}{\phi(a+\Delta)} 
&= \lim_{a \to \infty} \frac{\phi(a)}{\phi(a+\Delta)} \\
&= \lim_{a \to \infty} \frac{\phi'(a)}{\phi'(a+\Delta)} \\
&= \lim_{a \to \infty} \frac{e^{-a}}{1 + e^{-a}}\frac{1 + e^{-a-\Delta}}{e^{-a-\Delta}} \\
&= e^{\Delta} = e^{|b-a|},
\]
where the second equality follows from L'Hospital's rule. 
Similarly, if $\Delta < 0$,
\[
\sup_{a} \frac{\phi(a)}{\phi(a+\Delta)} 
&= \lim_{a \to -\infty} \frac{e^{-a}}{1 + e^{-a}}\frac{1 + e^{-a-\Delta}}{e^{-a-\Delta}} \\
&= \lim_{a \to -\infty} e^{\Delta}\frac{e^{-a}}{e^{-a-\Delta}} \\
&= 1 \le e^{|b-a|},
\]
where in this case we have used L'Hospital's rule twice. 
\eprf
\bnlem \label{lem:phi-convex}
The function $\phi(s)$ is convex. 
\enlem
\bprf
A straightforward calculation shows that $\phi''(s) = \frac{e^{s}}{(1+e^{s})^{2}} > 0$. 
\eprf

\bnlem \label{lem:expected-phi-ratio-bound}
For a random vector $Z \in \reals^{D}$ with finite mean $\bZ = \EE[Z]$ and a 
fixed vectors $V, \theta^{*} \in \reals^{D}$,
\[
\inf_{\theta \in \ball_{R}} \EE\left[\frac{\phi(Z \cdot (\theta + \theta^{*}))}{\phi(V \cdot (\theta + \theta^{*}))}\right] \ge e^{-R\|\bZ - V\|_{2} - |(\bZ - V)\cdot \theta^{*}|}. 
\]
\enlem
\bprf
Using \cref{lem:phi-ratio-bound,lem:phi-convex}, Jensen's inequality, and the triangle inequality, we have
\[
\inf_{\theta \in \ball_{R}} \EE\left[\frac{\phi(Z \cdot (\theta + \theta^{*}))}{\phi(V \cdot (\theta + \theta^{*}))}\right]
&\ge \inf_{\theta \in \ball_{R}} \frac{\phi(\EE[Z] \cdot (\theta + \theta^{*}))}{\phi(V \cdot (\theta + \theta^{*}))} \\
&\ge \inf_{\theta \in \ball_{R}} e^{-|(\bZ - V)\cdot (\theta + \theta^{*})|} \\
&\ge \inf_{\theta \in \ball_{R}} e^{-|(\bZ - V)\cdot\theta| - |(\bZ - V)\cdot\theta^{*})|} \\
&= e^{-R\|\bZ - V\|_{2} - |(\bZ - V)\cdot\theta^{*}|}.
\]
\eprf

We now prove the following generalization of \cref{lem:sensitivity-upper-bound}
\bnlem \label{lem:generalized-sensitivity-upper-bound}
For any $k$-clustering $\mcQ$, $\theta^{*} \in \reals^{d}$, and $R > 0$,
\[
\sigma_{n}(\theta^{*} + \ball_{R}) 
\le m_n \defined \left\lceil\frac{N}{1 + \sum_{i=1}^{k}|G_{i}^{(-n)}|e^{-R\|\bZ_{G,i}^{(-n)} - Z_{n}\|_{2} - |(\bZ_{G,i}^{(-n)} - Z_{n})\cdot\theta^{*}|}}\right\rceil.
\]
Furthermore, $m_{n}$ can be calculated in $O(k)$ time. 
\enlem
\bprf
Straightforward manipulations followed by an application of \cref{lem:expected-phi-ratio-bound} yield
\[
\sigma_{n}(\theta^{*} + \ball_{R})^{-1}
&= \inf_{\theta \in \ball_{R}} \frac{1}{N}\sum_{n'=1}^{N} \frac{\phi(Z_{n'} \cdot (\theta + \theta^{*}))}{\phi(Z_{n} \cdot (\theta + \theta^{*}))} \\
&= \inf_{\theta \in \ball_{R}} \frac{1}{N}\left[1 + \sum_{i=1}^{k}\sum_{Z' \in G_{i}^{(-n)}} \frac{\phi(Z' \cdot (\theta + \theta^{*}))}{\phi(Z_{n} \cdot (\theta + \theta^{*}))} \right] \\
&= \inf_{\theta \in \ball_{R}} \frac{1}{N}\left[1 + \sum_{i=1}^{k}|G_{i}^{(-n)}|\,\EE\left[\frac{\phi(Z_{G,i}^{(-n)} \cdot (\theta + \theta^{*}))}{\phi(Z_{n} \cdot (\theta + \theta^{*}))} \right]\right] \\
&\ge \frac{1}{N}\left[1 + \sum_{i=1}^{k}|G_{i}^{(-n)}|e^{-R\|\bZ_{G,i}^{(-n)} - Z_{n}\|_{2}  - |(\bZ_{G,i}^{(-n)} - Z_{n})\cdot\theta^{*}|}\right].
\]
To see that the bound can be calculated in $O(k)$ time, first note that
the cluster $i_{n}$ to which $Z_{n}$ belongs can be found in $O(k)$ time while
$\bZ_{G, i_{n}}^{(-n)}$ can be calculated in $O(1)$ time.
For $i \ne i_{n}$, $G_{i}^{(-n)} = G_{i}$, so $\bZ_{G, i}^{(-n)}$ is just the mean of cluster $i$, and no extra
computation is required.
Finally, computing the sum takes $O(k)$ time. 
\eprf

In order to obtain an algorithm for generating coresets for logistic regression, we 
require a bound on the dimension of  the range space constructed from the examples and 
logistic regression likelihood. 
\bnprop \label{prop:dimension-bound}
The set of functions $\mcF = \{ f_{n}(\theta) = \phi(Z_{n} \cdot \theta) \given n \in [N]\}$
satisfies $\dim(\mcF) \le D+1$. 
\enprop
\bprf
For all $F \subseteq \mcF$,
\[
|\{F \cap R \given R \in \ranges(\mcF)\}|
&= |\{\range(F, \theta, a) \given \theta \in \Theta, a \ge 0\}|,
\]
where $\range(F, \theta, a) \defined \{f_{n} \in \mcF \given f_{n}(\theta) \le a \}$. 
But, since $\phi$ is invertible and monotonic,
\[
\{f_{n} \in \mcF \given f_{n}(\theta) \le a \}
&= \{f_{n} \in \mcF \given \phi(Z_{n} \cdot \theta) \le a \} \\
&= \{f_{n} \in \mcF \given Z_{n} \cdot \theta \le \phi^{-1}(a) \},
\]
which is exactly a set of points shattered by the hyperplane classifier 
$Z \mapsto \sgn(Z \cdot \theta - b)$, with $b \defined \phi^{-1}(a)$. 
Since the VC dimension of the hyperplane concept class is $D+1$, 
it follows that~\citep[Lemmas 3.1 and 3.2]{Kearns:1994}
\[
|\{\range(F, \theta, a) \given \theta \in \Theta, a \ge 0\}|
&\le \sum_{j=0}^{D+1}{|F| \choose j} 
\le \sum_{j=0}^{D+1} \frac{|F|^{j}}{j!}  \\
&\le \sum_{j=0}^{D+1} {D+1 \choose j} |F|^{j}  
= (|F|+1)^{D+1}. 
\]
\eprf

\begin{proofof}{\cref{thm:coreset-algorithm}}
Combine \cref{thm:making-coresets,lem:sensitivity-upper-bound,prop:dimension-bound}. 
The algorithm has overall complexity $O(Nk)$ since it requires $O(Nk)$ time to calculate the sensitivities
by \cref{lem:sensitivity-upper-bound} and $O(N)$ time to sample the coreset.
\end{proofof}

\section{Sensitivity Lower Bounds}

\bnlem \label{lem:almost-equal-bad-vectors}
Let $V_{1},\dots,V_{K} \in \reals^{D-1}$ be unit vectors such that for some $\eps > 0$, for 
all $k \ne k$', $V_{k} \cdot V_{k'} \le 1 - \eps$. 
Then for $0 < \delta < \sqrt{1/2}$, there exist unit vectors $Z_{1},\dots,Z_{K} \in \reals^{D}$ such that
\bitems
\item for $k \ne k'$, $Z_{k} \cdot Z_{k'} \ge 1 - 2\delta^{2} > 0$
\item for $k = 1,\dots,K$ and $\alpha > 0$, there exists $\theta_{k} \in \reals^{D}$ such that 
$\|\theta\|_{2} \le \sqrt{2}\delta\alpha$, 
$\theta_{k} \cdot Z_{k} = -\frac{\alpha\eps\delta^{2}}{2}$ and for $k \ne k$, 
$\theta_{k} \cdot Z_{k'} \ge \frac{\alpha\eps\delta^{2}}{2}$. 
\eitems
\enlem
\bprf
Let $Z_{k}$ be defined such that $Z_{ki} = \delta V_{ki}$ for $i=1,\dots,D-1$ and $Z_{kD} = \sqrt{1-\delta^{2}}$.
Thus, $\|Z_{k}\|_{2} = 1$ and for $k \ne k'$, 
\[
Z_{k} \cdot Z_{k'} = \delta^{2}V_{k} \cdot V_{k'} + 1 - \delta^{2} \ge 1 - 2\delta^{2}
\]
since $V_{k} \cdot V_{k'} \ge -1$. 
Let $\theta_{k}$ be such that $\theta_{ki} = -\alpha\delta V_{ki}$ for $i=1,\dots,D-1$ and 
$\theta_{kd} = \frac{\alpha\delta^{2}(1-\eps/2)}{\sqrt{1-\delta^{2}}}$. 
Hence,
\[
\theta_{k} \cdot \theta_{k} &= \alpha^{2}\delta^{2}\left(V_{k} \cdot V_{k} + \frac{(1-\eps/2)^{2}\delta^{2}}{1-\delta^{2}}\right) \le 2\alpha^{2}\delta^{2} \\
\theta_{k} \cdot Z_{k} &= \alpha(-\delta^{2}V_{k} \cdot V_{k} + \delta^{2}(1-\eps/2)) = -\frac{\alpha\eps\delta^{2}}{2},
\]
and for $k' \ne k$,
\[
\theta_{k} \cdot Z_{k'} = \alpha(-\delta^{2}V_{k} \cdot V_{k'} + \delta^{2}(1-\eps/2)) \ge \alpha \delta^{2}(-1 + \eps + 1 - \eps/2) = \frac{\alpha\eps\delta^{2}}{2}.
\]
\eprf

\bnprop \label{prop:max-sensitivity-vectors}
Let $V_{1},\dots,V_{K} \in \reals^{D-1}$ be unit vectors such that for some $\eps > 0$, for 
all $k \ne k$', $V_{k} \cdot V_{k'} \le 1 - \eps$.
Then for any $0 < \eps' < 1$, there exist unit vectors 
$Z_{1},\dots,Z_{K} \in \reals^{D}$ such that for $k,k'$, $Z_{k} \cdot Z_{k'} \ge 1 - \eps'$ 
but for any $R > 0$,
\[
\sigma_{k}(\ball_{R}) \ge \frac{K}{1 + (K-1)e^{-R\eps\sqrt{\eps'}/4}},
\]
and hence $\sigma_{k}(\reals^{D}) = K$. 
\enprop
\bprf
Let $Z_{1},\dots,Z_{K} \in \reals^{D}$ be as in \cref{lem:almost-equal-bad-vectors} with 
$\delta$ such that $\delta^{2} = \eps'/2$. 
Since for $s \ge 0$, $\phi(s)/\phi(-s) \le e^{-s}$, conclude that, choosing $\alpha$
such that $\sqrt{2}\alpha\delta = R$, we have
\[
\sigma_{n}(\ball_{R}) 
&= \sup_{\theta \in \ball_{R}} \frac{K\,\phi(Z_{k} \cdot \theta)}{\sum_{k'=1}^{K} \phi(Z_{k'} \cdot \theta)} \\
&\ge \frac{K\,\phi(-\alpha\eps\delta^{2}/2)}{\phi(-\alpha\eps\delta^{2}/2) + (K-1)\phi(\alpha\eps\delta^{2}/2)} \\
&\ge \frac{K}{1 + (K-1)e^{-\alpha\eps\delta^{2}/2}} \\
&= \frac{K}{1 + (K-1)e^{-R\eps\sqrt{\eps'}/4}}.
\]
\eprf

\begin{proofof}{\cref{thm:sensitivity-lower-bound}}
Choose $V_{1},\dots,V_{N} \in \reals^{D-1}$ to be any $N$ distinct unit vectors.
Apply \cref{prop:max-sensitivity-vectors} with $K = N$ and 
$\eps = 1 - \max_{n \ne n'} V_{n} \cdot V_{n'} > 0$. 
\end{proofof}

\begin{proofof}{\cref{prop:epsilon-behavior}}
First note that if $V$ is uniformly distributed on $\sphere^{D}$, then the
distribution of $V \cdot V'$ does not depend on the distribution of $V'$ 
since $V \cdot V'$ and  $V \cdot V''$ are equal in distribution for all $V', V'' \in \sphere^{D}$.
Thus it suffices to take $V_{1}' = 1$ and $V_{i}' = 0$ for all $i=2,\dots,D$.
Hence the distribution of $V \cdot V'$ is equal to the distribution of $V_{1}$. 
The CDF of $V_{1}$ is easily seen to be proportional to the surface area (SA) 
of $C_{s} \defined \{ v \in \sphere^{D} \given v_{1} \le s \}$. 
That is, $\Pr[V_{1} \le s] = \text{SA}(C_{s})/\text{SA}(C_{1})$. 
Let $U \dist \distBeta(\frac{D-1}{2}, \frac{1}{2})$, and let $B(a,b)$ be the beta function.
It follows from \citep[Eq. 1]{Li:2011}, that by setting $s = 1 - \eps$
with $\eps \in [0,1/2]$,
\[
\Pr[V_{1} \ge 1 - \eps] 
&= \frac{1}{2} \Pr[-\sqrt{1 - U} \le \eps - 1] \\
&= \frac{1}{2} \Pr[U \le 2\eps - \eps^{2}] \\
&= \frac{1}{2 B(\frac{D-1}{2}, \frac{1}{2})} \int_{0}^{2\eps - \eps^{2}} t^{(D-3)/2}(1-t)^{-1/2}\,\dee t \\
&\le \frac{1}{2 B(\frac{D-1}{2}, \frac{1}{2})}(1-\eps)^{-1} \int_{0}^{2\eps - \eps^{2}} t^{(D-3)/2}\,\dee t \\
&= \frac{1}{(D-1) B(\frac{D-1}{2}, \frac{1}{2})}\frac{(2-\eps)^{(D-1)/2}}{1-\eps} \eps^{(D-1)/2} \\
&\le \frac{2^{(D+1)/2}}{(D-1) B(\frac{D-1}{2}, \frac{1}{2})} \eps^{(D-1)/2}.
\]
Applying a union bound over the ${D \choose 2}$ distinct vector pairs completes the proof.
\end{proofof}

\bnlem[{Hoeffding's inequality~\citep[Theorem 2.8]{Boucheron:2013}}] \label{lem:hoeffding}
Let $A_{k}$ be zero-mean, independent random variables with $A_{k} \in [-a,a]$. 
Then for any $t > 0$, 
\[
\Pr\left(\sum_{k=1}^{K}A_{k} \ge t\right) \le e^{-\frac{t^{2}}{2a^{2}K}}.
\]
\enlem
\begin{proofof}{\cref{prop:exponentially-many-distant-vectors}}
We say that unit vectors $V$ and $V'$ are \emph{$(1 - \eps)$-orthogonal} if $|V \cdot V'| \le 1-\eps$. 
Clearly $\|V_{n}\|_{2} = 1$.
For $n \ne n'$, by Hoeffding's inequality $\Pr(|V_{n} \cdot V_{n'}| \ge 1 - \eps) \le 2e^{-(1-\eps)^{2}D/2}$. 
Applying a union bound to all ${K \choose 2}$ pairs of vectors, the probability that any pair is
not $(1-\eps)$-orthogonal is at most
\[
2{K \choose 2} e^{-(1-\eps)^{2}D/2} \le \frac{1}{2}. 
\]
Thus, with probability at least $\half$, $V_{1},\dots,V_{N}$ are pairwise $(1-\eps)$-orthogonal. 
\end{proofof}

\begin{proofof}{\cref{prop:matching-sensitivity-bounds}}
The data from \cref{thm:sensitivity-lower-bound} satisfies $Z_{n} \cdot Z_{n'} \ge 1 - \eps'$,
so for $n \ne n'$,
\[
\|Z_{n} - Z_{n'}\|_{2}^{2} = 2 - 2Z_{n} \cdot Z_{n'} \le 2\eps'.
\]
Applying \cref{lem:sensitivity-upper-bound} with the clustering $\mcQ = \{Z_{1},\dots,Z_{N}\}$
and combining it with the lower bound in \cref{thm:sensitivity-lower-bound} yields the result. 
\end{proofof}

\section{A Priori Expected Sensitivity Upper Bounds}

\begin{proofof}{\cref{prop:mixture-upper-bound}}
First, fix the number of datapoints $N \in \mathbb{N}$.
Since $X_n$ are generated from a mixture, let $L_n$ denote the integer mixture component from which $X_n$ was generated,
let $C_i$ be the set of integers $1\leq j\leq N$ with $j\neq n$ and $L_j = i$, and let $C = (C_i)_{i=1}^\infty$.
Note that with this definition, $|G_i^{(-n)}| = |C_i|$.
 Using Jensen's inequality and the upper bound from \cref{lem:sensitivity-upper-bound} 
with the clustering induced by the label sequence, 
\[
\EE\left[\sigma_n\left(\ball_R\right)\right] \leq \EE\left[m_n\right] &= N\EE\left[\frac{1}{1 + \sum_i|C_i|e^{-R\|\bZ_{G,i}^{(-n)} - Z_{n}\|_{2}}} 
\right]\\
&= N\EE\left[\EE\left[\frac{1}{1 + \sum_i|C_i|e^{-R\|\bZ_{G,i}^{(-n)} - Z_{n}\|_{2}}} \given C\right]\right]\\
&\leq N\EE\left[\frac{1}{1 + \sum_i|C_i|e^{-R\EE\left[\|\bZ_{G,i}^{(-n)} - Z_{n}\|_{2}\given C\right]}} \right].
\]
Using Jensen's inequality again and conditioning on the labels $Y = (Y_n)_{n=1}^N$ and indicator $L_n$,
\[
\EE\left[\|\bar Z_{G, i}^{(-n)} - Z_n\|_2 \given C \right] &\leq \sqrt{\EE\left[\|\bar Z_{G, i}^{(-n)} - Z_n\|^2_2 \given C\right]}\\
&= \sqrt{\EE\left[\EE\left[\|\bar Z_{G, i}^{(-n)} - Z_n\|^2_2 \given C, L_n, Y\right] \given C\right]}.
\]
For fixed labels $Y$ and clustering $C$, $L_n$, the linear combination in the expectation is multivariate normal with
\[
\bar Z_{G, i}^{(-n)} - Z_n \dist \distNorm\left( \frac{1}{\left|C_i\right|}\left(\sum_{j \in C_i} Y_j\right)\mu_i - Y_n\mu'_n, \frac{1}{\left|C_i\right|}\Sigma_i + \Sigma'_n\right),
\]
where $\mu'_n, \Sigma'_n$ are the mean and covariance of the mixture component that generated $X_n$.
Further, for any multivariate normal random vector $W\in\mathbb{R}^d$,
\[
\EE\left[W^TW\right] &= \sum_{m=1}^d \EE\left[W_m^2\right] = \sum_{m=1}^d \var\left[W_m\right] + \EE\left[W_m\right]^2,
\]
so
\[
&\EE\left[\|\bar Z_{G, i}^{(-n)} - Z_n\|^2_2 \given L_n, C, Y\right] \\
=& 
\tr\left[\frac{1}{\left|C_i\right|}\Sigma_i + \Sigma'_n\right] 
+
 \left(\frac{\sum_{j\in C_i} Y_j}{\left|C_i\right|}\right)^2\mu_i^T\mu_i
-2Y_n\left(\frac{\sum_{j\in C_i} Y_j}{\left|C_i\right|}\right)\mu_i^T\mu'_n
+{\mu_n'}^{T}\mu'_n.
\]
Exploiting the \iid-ness of $Y_j$ for $j\in C_i$ given $C$, defining $\by_j = \EE\left[Y_i | L_i = j\right]$, and noting that $X_n$ is sampled from the mixture model,
\[
& \EE\left[\EE\left[\|\bar Z_{G, i}^{(-n)} - Z_n\|^2_2 \given L_n, C, Y\right] \given C\right] \\
=& \sum_j \pi_j\left(\tr\left[\frac{1}{\left|C_i\right|}\Sigma_i + \Sigma_j\right] 
+
 \frac{\left|C_i\right|\by_i^2 + 1-\by_i^2}{\left|C_i\right|}\mu_i^T\mu_i
-2\by_j\by_i\mu_i^T\mu_j
+\mu_j^T\mu_j\right)\\
=& \sum_j \pi_j\left(\frac{\tr\left[\Sigma_i\right] + \left(1-\by_i^2\right)\mu_i^T\mu_i}{\left|C_i\right|} + \tr\left[\Sigma_j\right] 
+
 \by_i^2\mu_i^T\mu_i
-2\by_j\by_i\mu_i^T\mu_j
+\mu_j^T\mu_j\right)\\
=& A_i \left|C_i\right|^{-1} + B_{in},
\]
where $A_i$ and $B_i$ are positive constants
\[
A_i &= \tr\left[\Sigma_i\right] + \left(1-\by_i^2\right)\mu_i^T\mu_i\\
B_i &= \sum_j \pi_j \left(\tr\left[\Sigma_j\right] 
+
 \by_i^2\mu_i^T\mu_i
-2\by_i\by_j\mu_i^T\mu_j
+\mu_j^T\mu_j\right).
\]
Therefore, with $0^{-1}$ defined to be $+\infty$,
\[
 \EE\left[m_n\right]&\leq N\EE\left[\frac{1}{1 + \sum_i|C_i|e^{-R\sqrt{A_i \left|C_i\right|^{-1} + B_{i}}}}\right].
\]
As $N\to\infty$, we expect the values of $|C_i|/N$ to concentrate around $\pi_i$. To get a finite sample bound 
using this intuition, we split the expectation into two conditional expectations: one where all
$|C_i|/N$ are not too far from $\pi_i$, and one where they may be.
Define $g : \mathbb{R}_+^\infty \to \mathbb{R}_+$ as
\[
g(x) =  \frac{1}{1 + \sum_{i} x_i e^{-R\sqrt{A_i x_i^{-1} + B_{i}}}},
\]
$\pi = (\pi_1, \pi_2, \dots)$, $\epsilon = (\epsilon_1, \epsilon_2, \dots)$ with $\epsilon_i > 0$,
and $\eta_{i} = \max(\pi_{i}-\epsilon_{i}, 0)$. Then
\[
  \EE\left[m_n\right] &\leq N\Pr\left(\forall i, \frac{|C_i|}{N} \geq \eta_i \right) g(N\eta)
+ 
N\Pr\left(\exists i: \frac{|C_i|}{N} < \eta_i\right) \\
&=Ng(N\eta)  + N\Pr\left(\exists i : \frac{|C_i|}{N} < \eta_i\right)\left(1 - g(N\eta)\right).
\]
Using the union bound, noting that $1-g(N\eta) \leq 1$, and then using Hoeffding's inequality yields
\[
\EE\left[m_n\right]
&\leq Ng(N\eta) + N\sum_{i}\Pr\left(\frac{|C_i|}{N} < \eta_i\right)\\
&\leq Ng(N\eta) + N\sum_{i : \pi_i > \epsilon_i}\Pr\left(\frac{|C_i|}{N} - \pi_i < -\epsilon_i\right)\\
&\leq Ng(N\eta) + N\sum_{i : \pi_i > \epsilon_i}e^{-2 N \epsilon_i^2}\\
&= \frac{1}{N^{-1} + \sum_{i} \eta_i e^{-R\sqrt{A_iN^{-1}\eta_i^{-1} + B_{i}}}}
 + \sum_{i : \pi_i > \epsilon_i}Ne^{-2 N \epsilon_i^2}.
\]
We are free to pick $\epsilon$ as a function of $\pi$ and $N$. 
Let $\epsilon = N^{-r}$ for any $0 < r < 1/2$. Note that 
this means $\eta_i = \max(\pi_i-N^{-r}, 0)$. Then
\[
\EE\left[m_n\right]
&= \frac{1}{N^{-1} + \sum_{i} \eta_i e^{-R\sqrt{A_iN^{-1}\eta_i^{-1} + B_{i}}}}
 + \sum_{i : \eta_i > 0}Ne^{-2 N^{1-2r}}.
\]
It is easy to see that the first term converges to 
$\left(\sum_{i} \pi_ie^{-R\sqrt{B_{i}}}\right)^{-1}$ by a simple asymptotic analysis. 
To show the second term converges to 0, note that for all $N$,
\[
\sum_i \pi_i &= \sum_{i : \pi_i > N^{-r}} \pi_i + \sum_{i : \pi_i \leq N^{-r}}\pi_i \\
&\geq \sum_{i : \pi_i > N^{-r}} \pi_i \\
&\geq \sum_{i : \pi_i > N^{-r}} N^{-r} \\
&= \left| \left\{i : \pi_i > N^{-r}\right\}\right| N^{-r}.
\]
Since $\sum_i \pi_i = 1 < \infty$, $\left| \left\{i : \pi_i > N^{-r}\right\}\right| = O(N^{r})$.
Therefore there exists constants $C, M < \infty$ such that
\[
\left|\left\{i : \pi_i > N^{-r}\right\}\right| &\leq M + C N^r,
\]
and thus
\[
\sum_{i : \pi_i > N^{-r}}Ne^{-2 N^{1-2r}} &\leq N(M+CN^r) e^{-2 N^{1-2r}} \to 0, \qquad N \to \infty.
\]
Finally, since $\bbm_N = \frac{1}{N}\sum_{n=1}^N m_n$, we have $\EE\left[\bbm_N\right] = \EE\left[m_n\right]$, and the result follows.
\end{proofof}
\begin{proofof}{\cref{cor:gaussian-upper-bound}}
This is a direct result of \cref{prop:mixture-upper-bound} with $\pi_1 = 1$, $\pi_i= 0$ for $i\geq 2$.
\end{proofof}

\section{Further Experimental Details}

The datasets we used are summarized in \cref{tbl:datasets}.
We briefly discuss some implementation details of our experiments.

\textbf{Implementing \cref{alg:lr-coreset}.}
One time-consuming part of creating the coreset is calculating the
adjusted centers $\bZ_{G,i}^{(-n)}$.
We instead used the original centers $Q_{i}$. 
Since we use small $k$ values and $N$ in large, each cluster is large.
Thus, the difference between $\bZ_{G,i}^{(-n)}$ and $Q_{i}$ was negligible 
in practice, resulting at most a 1\% change in the sensitivity while
resulting in an order of magnitude speed-up in the algorithm. 
In order to speed up the clustering step, we selected a random subset
of the data of size $L = \min(1000k, 0.025N)$ and ran the \texttt{sklearn}
implementation of $k$-means++ to obtain $k$ cluster centers.
We then calculated the clustering and the normalized $k$-means score
$\mcI$ for the full dataset. 
Notice that $L$ is chosen to be independent of $N$ as $N$ becomes large
but is never more than a construct fraction of the full dataset when $N$ is small.%
\footnote{Note that we use data subsampling here \emph{only} to choose the
cluster centers. We still calculate sensitivity upper bounds across
the entire data set and thereby are still able to capture rare but
influential data patterns. Indeed, we expect influential data points
to be far from cluster centers chosen either with or without
subsampling, and we thereby expect to pick up these data points with
high probability during the coreset sampling procedure in \cref{alg:lr-coreset}.}
Thus, calculating a clustering only takes a small amount of time that is comparable
to the time required to run our implementation of \cref{alg:lr-coreset}. 

\textbf{Posterior Inference Procedure.}
We used the adaptive Metropolis-adjusted Langevin algorithm~\citep{Haario:2001,Roberts:1996},
where we adapted the overall step size and targeted an acceptance rate of 0.574~\citep{Roberts:2001}.
It $T$ iterations were used in total, adaptation was done for the first $T/2$ iterations while
the remaining iterations were used as approximate posterior samples. 
For the subsampling experiments, for a subsample size $M$, an approximate dataset $\tilde\mcD$ of size $M$ was 
obtained either using random sampling or \cref{alg:lr-coreset}.
The dataset $\tilde\mcD$ was then fixed for the full MCMC run. 

\begin{table}[t]
\caption{Datasets used for experiments}
\begin{center}
\begin{tabular}{l|c|c|c|c}
Name & $N$ & $D$ & positive examples & $k$ \\ 
\hline
Low-dimensional Synthetic Binary & 1M & 5 &  9.5\% & 4 \\
Higher-dimensional Synthetic Binary & 1M & 10 &  8.9\% & 4 \\
Synthetic Balanced Mixture & 1M & 10 &  50\% & 4 \\
Chemical Reactivity\tablefootnote{Dataset \texttt{ds1.100} from \url{http://komarix.org/ac/ds/}.} & 26,733 & 100 & 3\% & 6 \\
Webspam\tablefootnote{Available from \url{http://www.cc.gatech.edu/projects/doi/WebbSpamCorpus.html}} & 350K & 127 & 60\% & 6 \\
Cover type\tablefootnote{Dataset \texttt{covtype.binary} from \url{https://www.csie.ntu.edu.tw/~cjlin/libsvmtools/datasets/binary.html}.} & 581,012 & 54 & 51\% & 6
\end{tabular}
\end{center}
\label{tbl:datasets}
\end{table}